\documentclass[a4paper,12pt]{article}

\usepackage[centertags]{amsmath}
\usepackage{amsthm,amsfonts,amssymb,amsmath,empheq,epsfig,graphicx}
\usepackage{hyperref}
\usepackage{color}

\newcommand{\ag}{{\alpha_G}}             
\newcommand{\amp}{M}                     
\newcommand{\ampRid}{\mathfrak{M}}       
\newcommand{\bk}[1]{\langle #1 \rangle}  
\newcommand{\bs}{\boldsymbol}
\newcommand{\bt}{{\boldsymbol{b}}}
\newcommand{\dif}{\mathrm{d}}            
\newcommand{\eik}{\mathrm{eik}}          
\newcommand{\el}{{\mathrm{el}}}          
\newcommand{\eps}{\epsilon}              
\newcommand{\esp}[1]{\mathrm{e}^{#1}}    
\newcommand{\GW}{\mathrm{GW}}            
\newcommand{\hyp}[4]{\,_1\hspace{-1pt} F_2(#1;#2,#3;#4)} 
\renewcommand{\Im}{\mathrm{Im\,}}        
\newcommand{\ini}{{\mathrm{in}}}         
\newcommand{\Lbq}{\mathcal{L}}           
\newcommand{\lp}{l_P}                    
\newcommand{\ls}{l_s}                    
\newcommand{\om}{\omega}
\newcommand{\omE}{\frac{\om}{E}}
\newcommand{\ord}[1]{\mathcal{O}\left(#1\right)}
\newcommand{\Phic}{\Phi_{\mathrm{c}}}    
\newcommand{\pol}{\epsilon}              
\newcommand{\psir}{\psi_{r}}              
\newcommand{\Qt}{{\boldsymbol{Q}}}
\newcommand{\qt}{{\boldsymbol{q}}}
\renewcommand{\Re}{\mathrm{Re}}          
\newcommand{\regge}{\mathrm{Regge}}
\newcommand{\rt}{{\boldsymbol{r}}}       
\newcommand{\sphere}{\mathbb{S}}         
\newcommand{\sgn}{\epsilon}              
\newcommand{\soft}{\mathrm{soft}}
\newcommand{\tomE}{\textstyle{\frac{\om}{E}}}
\newcommand{\tfa}{{\cal M}}              
\newcommand{\Tht}{{\boldsymbol{\Theta}}}
\newcommand{\Thr}{{\boldsymbol{\Theta}_{s,r}}} 
\newcommand{\tht}{{\boldsymbol{\theta}}}
\newcommand{\ui}{\mathrm{i}}             
\newcommand{\vq}{\vec{q}}                
\newcommand{\xt}{{\boldsymbol{x}}}
\newcommand{\zt}{{\boldsymbol{z}}}
\newcommand{\zetat}{{\boldsymbol{\zeta}}}

\numberwithin{equation}{section}

\begin{document}

\begin{titlepage}

\begin{center}
  {\Large \bf  Infrared features of gravitational scattering\\[1ex]
    and radiation in the eikonal approach}
 
\vskip 1.0cm

Marcello~Ciafaloni
\footnote{Email: ciafaloni@fi.infn.it}
\\
  {\sl\small Dipartimento di Fisica, Universit\`a di Firenze}\\
  {\sl\small Via Sansone 1, 50019 Sesto Fiorentino, Italy}\\[5mm]
  Dimitri~Colferai
  \footnote{Email: colferai@fi.infn.it}
  \\
  {\sl\small Dipartimento di Fisica, Universit\`a di Firenze
    and INFN, Sezione di Firenze}\\
  {\sl\small Via Sansone 1, 50019 Sesto Fiorentino, Italy}
  \\[3mm]
  and
  \\[3mm]
  Gabriele~Veneziano
  \footnote{Email: gabriele.veneziano@cern.ch}
  \\
  {\sl\small Theory Department, CERN, CH-1211 Geneva 23, Switzerland }\\
  {\sl\small Coll\`ege de France, 11 place M. Berthelot, 75005 Paris, France }
  \\[5mm]

\end{center}

\begin{abstract}
Following a semi-classical eikonal approach --- justified at transplanckian
energies order by order in the deflection angle
$\Theta_s\sim\frac{4G\sqrt{s}}{b} \equiv \frac{2 R}{b}$ --- we investigate the
infrared features of gravitational scattering and radiation in four space-time
dimensions, and we illustrate the factorization and cancellation of the
infinite Coulomb phase for scattering and the eikonal resummation for
radiation. As a consequence, both the eikonal phase $2\delta(E,b)$ and the
gravitational-wave (GW) spectrum $\frac{\dif E^\GW}{\dif\omega}$ are free from
infrared problems in a frequency region extending from zero to (and possibly
beyond) $\om =1/R$. The infrared-singular behavior of $4$-D gravity leaves a
memory in the deep infrared region ($\om R \ll \omega b < 1$) of the spectrum.
At $\ord{\om b}$ we confirm the presence of
logarithmic enhancements of the form already pointed out by Sen and
collaborators on the basis of non leading corrections to soft-graviton
theorems. These, however, do not contribute to the unpolarized and/or
azimuthally-averaged flux. At $\ord{\om^2 b^2}$ we find instead a positive
logarithmically-enhanced correction to the total flux implying an unexpected
maximum of its spectrum at $\om b \sim 0.5$. At higher orders we find
subleading enhanced contributions as well, which can be resummed, and have the
interpretation of a finite rescattering Coulomb phase of emitted gravitons.
\end{abstract}

\vskip 2cm
Preprint: CERN-TH-2018-268
\end{titlepage}


\section{Introduction\label{s:intro}} 

The recent discovery of gravitational waves (GW) in black-hole and neutron-star
mergers~\cite{LIGO,TheLIGOScientific:2017qsa} has also revived interest in
gravitational phenomena at the level of elementary-particle processes. It has
also been argued~\cite{Damour:2017zjx} that progress in the latter domain would
provide useful inputs on the determination of parameters that enter the
effective-one-body (EOB) approach~\cite{Buonanno:1998gg,Damour:2016gwp} to GW
emission from coalescing binary systems.

In particle physics, gravitational scattering of light particles or strings at
extremely high (i.e.\ transplanckian) energies has been considered since the
late eighties~\cite{tHooft:1987rb,ACV87,Muzinich:1987in, GrMe87, ACV88} mainly
as a thought-experiment aimed at testing quantum-gravity theories at very high
energies, and/or short distance.%
\footnote{In particular, the emergence of an
effective generalized uncertainty principle (GUP) holding in string theory has
been pointed out~\cite{ACV89} (see also~\cite{Veneziano:1986zf,Gross88}).}

At such energies, $\sqrt{s}/2\equiv E > m_P\equiv\sqrt{\hbar/G}$, and we meet a
regime in which the effective gravitational coupling $\ag\equiv Gs/\hbar$ is
large. Since such a coupling basically occurs as an overall factor in the
effective action (in $\hbar$ units) this suggests the validity of a
semiclassical approximation. This eikonal approach to high-energy gravitational
scattering was developed further by Amati, Ciafaloni and Veneziano
(ACV)~\cite{ACV89,ACV90,ACV93,ACV07} in a series of papers by deriving, in
particular, higher order corrections to the eikonal function.

Another emerging property of transplanckian gravitational scattering is a sort
of ``anti-scaling'' law by which the higher the center-of-mass energy, the
softer the characteristic energy of the final particles. This property has been
seen both in the string-size-dominated regime~\cite{ACV88, Veneziano:2004er}
and in the bremsstrahlung process, both
classically~\cite{GrVe14,Spirin:2015wwa} and at the quantum
level~\cite{CCV15,CCCV15,CC16,Dvali:2014ila,Addazi:2016ksu}. It is basically
related to the fact that multiplicities of final quanta grow like $\ag$ i.e.\ 
with two powers of the center of mass energy. Of course such a feature fits
extremely well with the well known behavior of the Hawking
temperature~\cite{Hawking:1974sw} of a black-hole of gravitational radius
$R\equiv2G\sqrt{s}$, $T\sim\hbar/R\ll E$. Interestingly, such a softening of
the final state already occurs in regimes (such as collision at large impact
parameter $b\gg R$) that are not expected to lead to black hole formation. Our
study of gravitational bremsstrahlung will concentrate therefore exclusively on
the regime $\frac{\hbar \om}{\sqrt{s}} \ll \frac{\hbar \om}{m_P} \ll 1$. Note
that this does not prevent considering a wide range of frequencies all the way
from zero, to $1/b$, to $1/R$, or even higher.

More recently, the low-frequency gravitational bremsstrahlung spectrum has also
been investigated~\cite{LaSe18,SaSe18,ABV2} in connection with Weinberg's
soft-graviton theorem~\cite{We65} and its extension to subleading
orders~\cite{He:2014laa,Strominger:2014pwa,Schwab:2014xua,Bern:2014oka,Afkhami-Jeddi:2014fia,Bianchi:2014gla,Bern:2014vva,Sen:2017xjn,Sen:2017nim,Guerrieri:2017ujb}.
The possible emergence of large soft logarithms (in $D=4$) has been recently
emphasized in~\cite{LaSe18,SaSe18} as subleading contributions to soft theorems
and a possible source of memory effects. This approach, unlike the eikonal one,
is not limited to high energy or to small deflection angles, but only covers a
tiny region of frequencies (basically the one below $1/b \ll 1/R$). Thus
comparison of the two approaches is necessarily limited to the extreme lower
end of the $\om$ spectrum.

The purpose of the present paper is to illustrate the essentials of the eikonal
model just mentioned, and then to focus on the derivation of soft-graviton
features, in order to see whether they are affected by the $D=4$ infrared (IR)
singularity of the gravitational interaction. 

We should notice from start that, in our approach, we shall mostly refer to
scattering at fixed impact parameter $b$, rather than fixed momentum transfer
$Q$. In $b$-space the $S$-matrix exponentiates both the eikonal function
$\delta(b,E)$, which controls time-delay and deflection angle, and the
multi-graviton production amplitudes in the form of a coherent state
immediately connected to classical GW radiation.

An important goal of the paper is to show (sec.~\ref{s:ilr}) that the eikonal
resummation --- which is needed in order to cover sizeable deflection angles of
order $\Theta_E\equiv 2R/b$ (the Einstein deflection angle) --- is also able to
build up divergence-free amplitudes. That is true both for scattering (due to
the factorization in impact parameter space and to the cancellation~\cite{ACV90}
of the infinite Coulomb phase at that order) and for radiation (due to the
smoothing out of the single-exchange amplitude by $s$-channel iteration).

Given such a regular behavior of the resumed amplitude, the study of soft
limits is straightforward, and based on the simple form of the resummed
radiation amplitude in the classical limit given in secs.~\ref{s:ilr},
\ref{s:sob}. At leading level, the energy emission spectrum --- as already
discussed in~\cite{GrVe14,CCV15,CCCV15} --- shows a $\log(1/\om R)$ dependence
in the intermediate-frequency region $1/b \ll \om < 1/R$, before saturating at
the expected $\om$-independent zero-frequency limit~\cite{Sm77}. At subleading
level, the rescattering Coulomb phase shows up in its finite and exponentiated
form, generating a class of logs of relative order $[b\om\log(1/b\om)]^n$ in
the $b\om \ll 1$ limit, similar (if not identical) to those already
proposed in~\cite{LaSe18,SaSe18}.

With the aim of being as much as possible self-contained the rest of the paper
is organized as follows: In sec.~\ref{s:es} we recall some old results on the
eikonal approximation to high energy elastic gravitational scattering. In
sec.~\ref{s:ugea} we recall previous analysis of the single graviton emission
amplitude and, in particular, our unified description of both the very soft
(Weinberg) regime and not so soft (Lipatov) one. These results are then used in
sec.~\ref{s:iles} to recover in a simple way a previous result on the subleading
correction to the eikonal phase and deflection angle.  In sec.~\ref{s:ilr} we
present the basic starting point for our study of soft gravitational
bremsstrahlung in the form of an infrared-finite unitary $S$-matrix which
agrees, in the appropriate limit, with the classical calculation obtained
earlier by completely different techniques.  Sec.~\ref{s:sob} contains most of
the new results of this work both on the sub-leading correction to circularly
polarized spectra and on the sub-sub-leading positive, logarithmically enhanced,
corrections to the ZFL in the frequency region $\om b \ll 1$. We also show how
this regime connects smoothly with a logarithmically decreasing one in the
region $1/b < \om < 1/R$ leading to a peak in the flux around $\om b \sim 0.5$
(and roughly independent of $R$).  In sec.~\ref{s:disc} we discuss our results
and point to possible directions for future research.

\section{Elastic eikonal scattering: a reminder\label{s:es}} 

In this section we summarize the ideas and assumptions introduced
in~\cite{CCCV15} in order to understand the main ingredients that our eikonal
radiation picture is based upon.

Throughout this paper, as in~\cite{ACV07}, we will restrict our attention to
collisions in 4-dimensional space-time and in the point-particle (or quantum
field theory) limit.  Consider the elastic gravitational scattering
$p_1+p_2 \to p'_1 + p'_2$ of two ultrarelativistic particles, with external
momenta parametrized as%
\footnote{Boldface symbols denote transverse vectors.}
\begin{equation}\label{mompar}
  p_i = E_i (1, \Tht_i, \sqrt{1 - |\Tht_i|^2}) \, ,
\end{equation}
at center-of-mass energy $2E=\sqrt{s}\gg M_P$ and momentum transfer
$Q^\mu\equiv p_1^{\prime\mu}-p_1^\mu = p_2^\mu - p_2^{\prime\mu}$ with
transverse component $\Qt=E\Tht_s$; the 2-vectors
$\Tht_i=|\Tht_i|(\cos\phi_i,\sin\phi_i)$ describe both azimuth $\phi_i$ and
polar angles%
\footnote{Strictly speaking, if $\Theta_i$ denotes the standard polar angle,
  $|\Tht_i|=\sin(\Theta_i)$. In the small-angle kinematics we deal with,
  $|\Tht_i|\simeq\Theta_i$.\label{n:angles}}
$|\Tht_i|\ll 1$ of the corresponding 3-momentum with respect to the
longitudinal $z$-axis.

This regime is characterized by a strong effective coupling
$\alpha_G\equiv Gs/\hbar \gg 1$ and was argued by several
authors~\cite{tHooft:1987rb,Muzinich:1987in,ACV88,ACV90} to be described by an
all-order leading approximation which has a semiclassical effective metric
interpretation.  The leading result for the $S$-matrix $S(b,E)$ in
impact-parameter $\bt$ space has the eikonal form
\begin{equation}\label{eikform}
  S(b,E) = \exp[2\ui\delta_0(b,E)]\;, \qquad
  \delta_0(b,E) = \alpha_G \log\frac{L}{b} \;, \qquad b\equiv|\bt| \;,
\end{equation}
$L$ being a factorized --- and thus unobservable --- IR cutoff due to the 
infinite Coulomb phase~\cite{ACV88}.

Corrections to the leading form~\eqref{eikform} involve additional powers of
the Newton constant $G$ in two dimensionless combinations
\begin{equation}\label{adimComb}
  \frac{\hbar G}{b^2}=\frac{\lp^2}{b^2} \;, \qquad
  \frac{4G^2 s}{b^2}=\frac{R^2}{b^2} \sim\alpha_G\frac{\lp^2}{b^2}
  \gg \frac{\lp^2}{b^2} \,,
\end{equation}
$\lp\equiv\sqrt{\hbar G}$ being the Planck length.  Since $\alpha_G \gg1$ we can
neglect completely the first kind of corrections.  Furthermore, we can consider
the latter within a perturbative framework since the impact parameter $b$ is
much larger than the gravitational radius $R\equiv 2G\sqrt{s}$.

\begin{figure}[ht]
  \centering
  \includegraphics[width=0.6\linewidth]{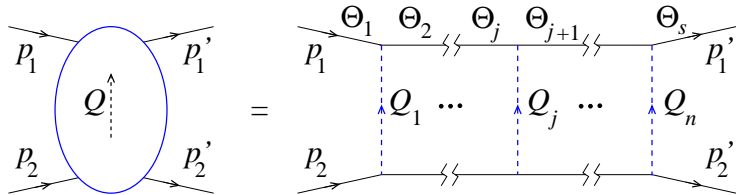}
  \caption{\it The scattering amplitude of two transplanckian particles (solid
    lines) in the eikonal approximation. Dashed lines represent (reggeized)
    graviton exchanges. The fast particles propagate on-shell throughout the
    whole eikonal chain. The angles $\Tht_j \simeq \sum_{i=1}^{j-1} \tht_i$
    denote the direction of particle 1 w.r.t.\ the $z$-axis along the scattering
    process.}
  \label{f:eikChain}
\end{figure}

In order to understand the scattering features implied by~\eqref{eikform} we can
compute the $\Qt$-space amplitude
\begin{equation}
 \frac{1}{s} \amp_\eik(s,\Qt^2)
  = 4 \int\dif^2\bt\;\esp{-\frac{\ui\bt\cdot\Qt}{\hbar}}
  \frac{\esp{2\ui\delta_0(b,E)}}{2\ui} 
  =\frac{8\pi\alpha_G}{\Qt^2}
  \left(\frac{4 \hbar^2}{\Qt^2 L^2}\right)^{-\ui\alpha_G}
  \frac{\Gamma(1-\ui\alpha_G)}{\Gamma(1+\ui\alpha_G)} \;, \label{QspAmp}
\end{equation}
where the last expression  is obtained strictly-speaking by extending
the $\bt$-integration up to small $|\bt|\lesssim R$~\cite{tHooft:1987rb}, where
corrections may be large. But it is soon realized that the $\bt$-integration
in~\eqref{QspAmp} is dominated by the saddle-point
\begin{equation}\label{saddlepoint}
  \Qt = E\Tht_s(\bt) = - E\frac{2R}{b}\hat{\bt} = - 2 \alpha_G
  \frac{\hbar}{b}\hat{\bt} \;,
\end{equation}
which leads to the same expression for the amplitude, apart from an irrelevant
$\Qt$-independent phase factor. The saddle-point momentum
transfer~\eqref{saddlepoint} comes from a large number $\bk{n}\sim\alpha_G$ of
graviton exchanges (fig.~\ref{f:eikChain}), corresponding to single-hit momentum
transfers $\bk{|\qt_j|}\simeq\hbar/b$ which are small, with very small
scattering angles $|\tht_j|$ of order $\theta_m\simeq\hbar/(bE)$. The overall
scattering angle --- though small for $b\gg R$ --- is much larger than
$\theta_m$ and is $|\Tht_s|=2R/b = 2\alpha_G\theta_m$, the Einstein deflection
angle.

In other words, every single hit is effectively described by the elastic
amplitude
\begin{equation}
  \amp_{\el}(\Qt_j) = \frac{\kappa^2 s^2}{\Qt_j^2}
  = \frac{\kappa^2 s^2}{E^2 \tht_j^2},
  \quad \left(\kappa^2 = \frac{8\pi G}{\hbar}\right) \; ,
\end{equation}
which is in turn directly connected to the phase shift $\delta_0$:%
\footnote{Here we use a cutoff regularization of IR $\Qt$'s, i.e.,
$|\Qt|>\hbar/L$ so as to recover the leading eikonal $\delta_0=\ag\log(L/b)$.}
\begin{equation}
  \delta_0(|\bt|,E) = \frac{1}{4s} \int \frac{\dif^2 \Qt}{(2\pi)^2}
  \esp{\frac{{\ui \Qt \cdot \bt}}{\hbar}} \amp_{\el}(\Qt)
  = \alpha_G \int \frac{\dif^2 \tht_s}{2\pi \tht_s^2}
  \esp{\frac{{\ui E \tht_s \cdot \bt}}{\hbar}} \; .
\end{equation}

The relatively soft nature of transplanckian scattering just mentioned is also
--- according to~\cite{ACV88} --- the basis for its validity in the
string-gravity framework.
Furthermore, the multiple-hit procedure can be generalized to multi-loop
contributions in which the amplitude, for each power of $G$, is enhanced by
additional powers of $s$, due to the dominance of $s$-channel iteration in
high-energy spin-2 exchange versus the $t$-channel one (which provides at most
additional powers of $\log s$). That is the mechanism by which the $S$-matrix
exponentiates an eikonal function (or operator) with the effective coupling
$\alpha_G\equiv Gs/\hbar$ and subleading contributions which are a power series
in $R^2/b^2$.

Both the scattering angle~\eqref{saddlepoint} (and the
$S$-matrix~\eqref{eikform}) can be interpreted from the metric point of
view~\cite{tHooft:1987rb} as the geodesic shift (and the quantum matching
condition) of a fast particle in the Aichelburg-Sexl (AS)
metric~\cite{AiSe70} of the other.

More directly, the associated metric emerges from the calculation~\cite{CC14} of
the longitudinal fields coupled to the incoming particles in the eikonal series,
which turn out to be
\begin{align}
  \frac14 h^{++} = h_{--} &= 2\pi R a_0(\xt)
  \delta\left(x^- -\pi R\sgn(x^+)a_0(b)\right) \;, \nonumber \\
  a_0(\xt) &= \frac1{2\pi}\log\frac{L^2}{\xt^2}\;, \qquad
  \delta_0(b,E) = \pi \alpha_G a_0(b) \;. \label{hpp}
\end{align}
Such shock-wave expressions yield two AS metrics for the fast particles, as well
as the corresponding time delay and trajectory shifts at leading level. When $b$
decreases towards $R\gg\ls$, corrections to the eikonal and to the
effective metric involving the $R^2/b^2$ parameter have to be included, 
as well as graviton radiation, to which we now turn.

\section{The unified single-graviton emission amplitude\label{s:ugea}} 

We start, in the ACV framework, from the irreducible (possibly
resummed~\cite{CC16}) eikonal, which in $D=4$ takes the form
\begin{equation}\label{deltabs}
  \delta(\bt,E)\equiv\ag\left[\log\frac{L}{R}+\Delta(\bt/R)  
  \right] =  \frac{ER}{\hbar} \left[\log\frac{L}{R}+\Delta(\bt/R)  
  \right] \;,
\end{equation}
that we split into an IR divergent ``Coulomb'' contribution regularized by the
cutoff $L$, and a finite part $\Delta$ which embodies the $\bt$ dependence. The
IR divergent Coulomb phase factorizes in front of the $S$-matrix~\cite{CCCV15}
and should cancel out in measurable quantities. The Fourier transform of
$\Delta(\bt)$ defines a ``potential'' $\tilde\Delta(\Qt)$ in transverse space.
In particular, the leading eikonal $\delta_0(\bt,E) = \ag\log(L/|\bt|)$
corresponds to $\tilde\Delta(\Qt)=1/\Qt^2 \times \Theta(\Qt^2-(\hbar/L)^2)$.

\begin{figure}[ht]
  \centering
  \includegraphics[width=0.5\linewidth]{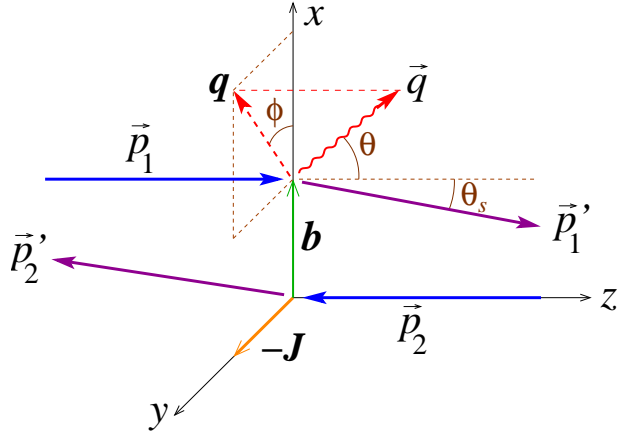}
  \caption{\it Center-of-mass view of the collision at impact parameter
      $\bt$ of particles 1 and 2 with associated emission of a graviton $q$. The
      polar angles $\Theta_s$ and $\theta$ are related to the 2D vectors
      $\Tht_s$ and $\tht$ as described in eq.~\eqref{mompar} and
      footnote~\ref{n:angles}.}
    \label{f:3DimpactParam}
\end{figure}

Consider now, at tree level, the emission of a graviton with energy $\hbar\om$
and transverse momentum $\qt=\hbar\om\tht$, $|\tht|$ being related to the polar
emission angle while $\phi_{\tht}$ is the azimuth in the transverse plane
(fig.~\ref{f:3DimpactParam}).
Keeping in mind that the condition $\hbar\om\ll E$ is always assumed in this
paper, we can still distinguish a ``Weinberg limit'' in which $|\qt| < |\Qt|$
for which the emission amplitude is given by Weinberg' external-line insertion
formula, and a ``Regge-Lipatov regime'' in which $|\qt| > |\Qt|$ so that
emission from the exchanged (and now effectively on shell) graviton has to be
added. Fortunately a single, simple expression~\cite{CCCV15,CC16} is able to
cope simultaneously with both regimes. Let us briefly discuss how.

\begin{figure}[ht]
  \centering
  \includegraphics[width=0.7\linewidth]{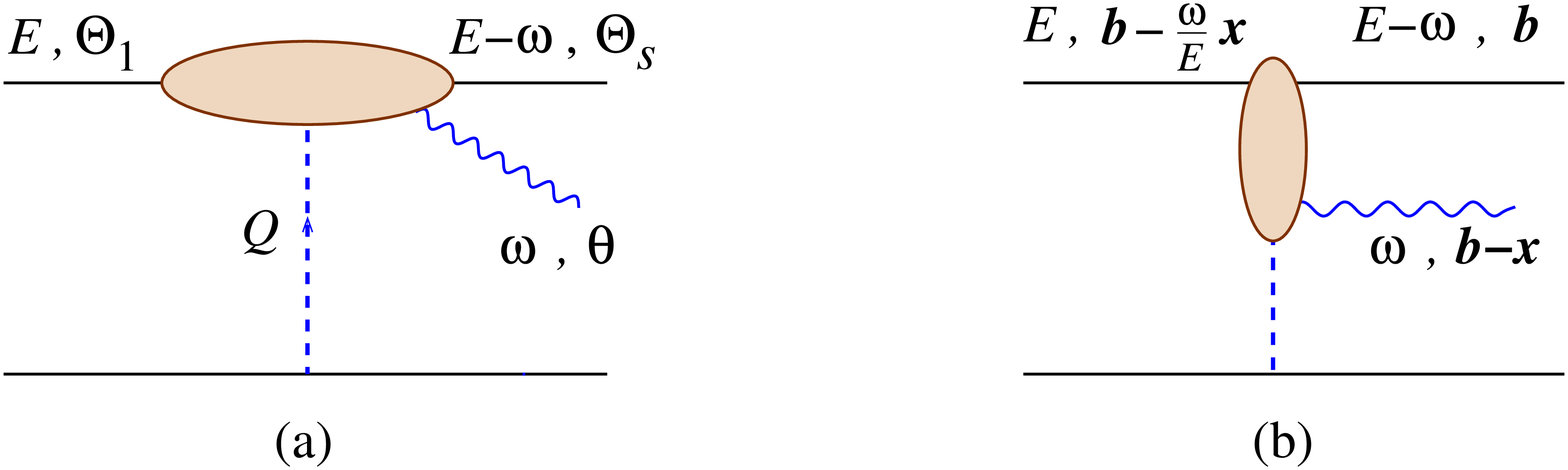}
  \caption{\it Single-exchange emission diagram in $\Qt$-space with deflection 
  angles {\rm(a)}, and its transverse-space counterpart with final-state
  variables $\bt$, $\xt$ and the shifted impact parameter $\bt-\omE\xt${}
	{\rm(b)}.}
    \label{f:upperInsertions}
\end{figure}

Weinberg's external insertion recipe factorizes in $\Qt$-space
(fig.~\ref{f:upperInsertions}a). This can be translated in $\bt$-space as
follows~\cite{CCCV15} (setting momentarily $\hbar=1$):
\begin{equation}\label{Msoft}
  \tfa^\soft_\lambda(\bt,E,\qt,\om) = \sqrt{\ag}\frac{R}{\pi}\int
  \frac{\dif^2\Qt}{2\pi}\;\tilde\Delta(\Qt)\esp{\ui\Qt\cdot\bt}
  \left[\frac{E}{\om}\,\frac12\left(
      \esp{-\ui\lambda(\phi_{\qt-\frac{\om}{E}\Qt} -\phi_\qt)}-1
      \right)\right] \;,
\end{equation}
where $\lambda=\pm2$ is the helicity of the emitted graviton, and the factor in
square brackets comes from the explicit computation of the Weinberg current on
helicity states. The latter are conveniently defined by the 
polarization tensors~\cite{CCV15,CCCV15}
\begin{align}
  \pol_\pm^\mu
  &= \frac1{\sqrt{2}}\left(\pol_{TT}^{\mu\nu}\pm\ui\pol_{LT}^{\mu\nu}\right)
  =\frac12 \left(\pol_T^\mu\pm\ui\pol_L^\mu\right){}
   \left(\pol_T^\nu\pm\ui\pol_L^\nu\right) \nonumber \\
  \pol_{TT}^{\mu\nu} &= \frac1{\sqrt{2}}
  \left(\pol_T^\mu\pol_T^\nu-\pol_L^\mu\pol_L^\nu\right) \;,\quad
 \pol_{LT}^{\mu\nu} = \frac1{\sqrt{2}}
 \left(\pol_L^\mu\pol_T^\nu+\pol_T^\mu\pol_L^\nu\right) \nonumber \\
 \pol_T^\mu &= \Big(0,-\eps_{ij}\frac{q_j}{|\qt|},0\Big) \;, \qquad
 \pol_L^\mu  = 
 \Big(\frac{q^3}{|\qt|},\bs{0},\frac{q^0}{|\qt|}\Big)\mp\frac{q^\mu}{|\qt|} \;,
 \label{pol}
\end{align}
where $\eps_{12}=1$ and the $-$ ($+$) sign in $\pol_L^\mu$ corresponds to 
graviton emission in the forward (backward) hemisphere in the small-angle 
kinematics.

We note that the phase difference in~\eqref{Msoft} can also be written in terms 
of deflection angles as
$\phi_{\qt-\frac{\om}{E}\Qt} -\phi_\qt = \phi_{\tht-\Tht_s}-\phi_\tht $ and can
be expressed by the integral representation
\begin{equation}\label{zRep}
  \esp{2\ui \phi_{\tht}} - \esp{2\ui \phi_{\tht'}}
  = -2 \int \frac{\dif^2\xt}{{2\pi x^*}^2} \left( \esp{\ui\om\xt \cdot \tht}
    - \esp{\ui\om\xt \cdot \tht'} \right) \;,
\end{equation}
where $x\equiv x_1+\ui x_2$, $x^*\equiv x_1-\ui x_2$ is the complex notation
for the transverse vector $\xt$ to be interpreted as the transverse distance
between the forward outgoing hard particle and the emitted graviton. In
addition, the Fourier transform~\eqref{Msoft} identifies $\bt$ as the
transverse distance between the two {\em outgoing} hard particles, so that
$\bt-\xt$ is the transverse coordinate of the emitted graviton w.r.t.\ the
backward hard particle (whose transverse position is essentially unaffected by
the forward emission process), as shown in fig.~\ref{f:upperInsertions}b. In
terms of such final-state variables, the impact parameter of the two incoming
hard particles amounts to $\bt_\ini=\bt-\omE\xt$. It is also interesting to
note that the classical orbital angular momentum
$(L^{13},L^{23})\simeq\sum_{p} E_{p}\rt_{p}$ is conserved in the process.

Inserting eq.~\eqref{zRep} into eq.~\eqref{Msoft}, it is straightforward to
perform the $\Qt$ integrals in terms of eikonal functions of linear combinations
of $\bt$ and $\xt$, thus yielding
\begin{equation}\label{softamp}
  \tfa^{\soft}_\lambda(\bt,E,\qt,\om) = -\sqrt{\ag}\frac{R}{\pi}
  \esp{\ui\lambda\phi_\qt}
  \int\frac{\dif^2\xt}{2\pi|\xt|^2\esp{\ui\lambda\phi_\xt}}\;
  \esp{\ui\qt\cdot\xt} \frac{E}{\om}
  \Big(\Delta\big(\bt-\tomE\xt\big)-\Delta(\bt)\Big)\;,
\end{equation}
which expresses the Weinberg insertions in $\bt$-space in terms of the eikonal 
functions with shifted impact parameter value $\bt-\omE\xt$ 
(fig.~\ref{f:upperInsertions}).

Furthermore, it was shown in~\cite{CCCV15} that the difference between the
Regge and soft amplitude in the overlapping soft-central region of phase space
is formally equal to (minus) the soft amplitude itself, provided one replaces
the scale parameter $E$ with $\om$. In other words, the unifying amplitude
$\tfa^{[1]}$ matching $\tfa^\soft$ and $\tfa^\regge$ in the corresponding
phase-space validity regions can be represented as%
\footnote{The superscript $^{[1]}$ indicates that we are still dealing with a
  single-exchange amplitude.}
\begin{equation}
 \tfa^{[1]}(\bt,E,\qt,\om) \simeq \tfa^\soft(\bt,E,\qt,\om) -
  \tfa^\soft(\bt;\om,\qt,\om) \;.
\end{equation}
In conclusion, the unified single-exchange amplitude reads
\begin{align}
  \tfa^{[1]}_\lambda(\bt,E,\qt,\om) &= -\sqrt{\ag}
  \frac{R}{\pi}\esp{\ui\lambda\phi_\qt}
  \int\frac{\dif^2\xt}{2\pi|\xt|^2\esp{\ui\lambda\phi_\xt}}\;
  \esp{\ui\qt\cdot\xt} \nonumber \\
  &\quad\times \left[
    \frac{E}{\om}\Big(\Delta\big(\bt-\tomE\xt\big)-\Delta(\bt)\Big) -
    \Big(\Delta(\bt-\xt)-\Delta(\bt)\Big)\right] \label{uniamp} \\
  &\simeq \sqrt{\ag}\frac{R}{\pi}\esp{\ui\lambda\phi_\qt}
  \int\frac{\dif^2\xt}{2\pi|\xt|^2\esp{\ui\lambda\phi_\xt}}\;
  \esp{\ui\qt\cdot\xt} \left[
 \Delta'(b)\, \xt\cdot\hat\bt +
    \Delta(\bt-\xt)-\Delta(\bt) \right] \;, \label{svilamp}
\end{align}
where, by considering an angular range $\theta=\ord{1/\om|\xt|}\gg 1/Eb$ we
have directly taken the $\om=0$ limit of the ``insertion function''
\begin{equation}\label{Phic}
  \Phic\Big(\frac{x}{b}\Big) \equiv \Phi(\bt,\xt,\om=0){}
   = -\Delta'(b)\, \xt\cdot\hat\bt + \Delta(\bt)-\Delta(\bt-\xt)
  = \Re\left( \frac{x}{b}+\log\big(1-\frac{x}{b}\big) \right) \;,
\end{equation}
which thereby acquires a classical meaning.

We notice that eq.~\eqref{uniamp} is directly expressed in terms of the eikonal
function $\ag\Delta(\bt)$ of eq.~\eqref{deltabs}, where the first (second) term
in square brackets is in correspondence with external (internal) insertions,
representing the Weinberg current (the high-energy correction). Furthermore,
the Weinberg part is proportional to the classical scattering angle
$\Theta_s=2R \Delta'(b)$ and produces the leading $1/\om$ behaviour of the
amplitude.

By then replacing~\eqref{Phic} into~\eqref{uniamp} we obtain the single-exchange
emission amplitude in the soft-based representation (e.g.\ for $\lambda=-2$)
\begin{equation}\label{Msb}
  \tfa^{[1]}(\bt,E;\qt,\om) = \sqrt{\ag}\frac{R}{2}\frac{q^*}{q}
  \int\dif^2\xt\; \esp{\ui\qt\cdot\xt} h_s(\bt,\xt) \;,
\end{equation}
where the soft field $h_s$ --- in the small-deflection regime described by the
leading eikonal --- has the expression
\begin{equation}\label{hsxxs}
  h_s(\bt,\xt) = -\frac{\Re\left(\frac{x}{b}
  +\log\big(1-\frac{x}{b}\big)\right)}{\pi^2 x^{*2}} \;.
\end{equation}

\section{Infrared logs in the elastic eikonal phase\label{s:iles}}

The long distance features of the Coulomb-like interaction mentioned before at
leading level $\sim\ag$, affect gravitational scattering at higher orders as
well. ACV~\cite{ACV90} provided a calculation of the next few orders in the
eikonal, which in our massless transplanckian scattering involve the parameters
$l_p^2/b^2$ and $R^2/b^2$ introduced before. Here we recall those results, and
we illustrate them in order to gain some better understanding of the role of the
IR singularity for graviton radiation as well.

\begin{figure}[hb]
  \centering
  \includegraphics[width=0.22\linewidth]{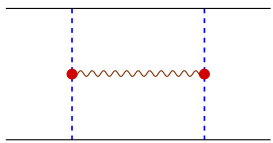}
  \caption{\it The H diagram providing the first subleading correction 
    to the eikonal phase.}
  \label{f:Hdiagram}
\end{figure}

Due to the exponentiation of the $S$-matrix in impact parameter space, we have
the second-order expansion
\begin{equation}\label{Sel}
  S_\el = \esp{2\ui\delta(b,E)} = \esp{2\ui(\delta_0+\delta_1+\delta_2+\cdots)}
    =1+2\ui(a^{(0)}+a^{(1)}+a^{(2)}+\cdots)
\end{equation}
where fixed-order amplitudes $a^{(n)}$ are related to the eikonal coefficients
$\delta^{(n)}$ as follows:
\begin{align}
  a^{(0)} &= \delta_0 = \frac{Gs}{\hbar}\log\frac{L}{b}
   = \frac{Gs}{\hbar}\left(\log\frac{L}{R} + \log\frac{R}{b}\right) \label{a0}\\
  a^{(1)} &= \ui\delta_0^2+\delta_1\;, \label{a1} \\
  a^{(2)} &= -\frac23\delta_0^3 +2 \ui\delta_0 \delta_1 + \delta_2 \;.\label{a2}
\end{align}
We noticed already that the cutoff dependence in $a_0$ is additive in impact
parameter space, and is thus factorizable in the $S$-matrix as a pure overall
phase, which is, by itself, unobservable.
But we want to look at higher orders also, and in particular at order
$\ag R^2/b^2$. For such terms the ACV method was to compute the imaginary parts
of the measurable parameters $\delta_1,\,\delta_2$ as required by unitarity
diagrams, and to derive the real parts by analyticity and asymptotic behaviour
arguments. For pure gravity they set
\begin{align}
 \Im\delta_1 &= 0 = \Im a^{(1)} -\delta_0^2 \label{imd1} \\
 \Im\delta_2 &= \Im a^{(H)} \;,\qquad \text{yielding in total} \label{imd2}\\
 \Im a^{(2)} &= 2\delta_0\delta_1 + \Im a^{(H)} \label{ima2}
\end{align}
In eq.~\eqref{ima2} the first term represents the 2-body discontinuity and the
second one the $2\to3$ contribution to $\Im\delta_2$, due to graviton radiation
in the central region, as embodied in the H-diagram (fig.~\ref{f:Hdiagram}). At
this point, ACV looked for analytic functions of the Mandelstam variables having
the correct discontinuities and asymptotic behaviours of $\delta_1$ and
$\delta_2$, so as to determine both.

At one-loop level, starting from eq.~\eqref{imd1}, they found only one 
analytic structure, yielding
\begin{equation}\label{a1v}
  a^{(1)} = \ui\delta_0^2+\delta_1 = \left(\ui+\frac{3}{\pi}\log s
  \frac{\hbar^2 \nabla_\bt^2}{s}\right) \delta_0^2(b,E)
\end{equation}
and thus determining in this way the one-loop coefficient
\begin{equation}\label{delta1}
  \delta_1 = \frac{6}{\pi}\frac{Gs}{\hbar}\frac{l_p^2}{b^2}\log s
  = \frac{6}{\pi}\frac{G^2 s}{b^2}\log s \qquad\text{(pure gravity)}\;.
\end{equation}

The above result for $\delta_1$ is consistent with what has been obtained
starting from supergravity calculations~\cite{Li88} after
subtracting~\cite{ABC90} the gravitino contribution.  We also checked that it
agrees with more recent estimates%
\footnote{We are grateful to Pierre Vanhove for having brought this reference to our attention.}
\cite{Dunbar:1994bn}.  We are not aware, instead, of any independent calculation
of $\delta_2$.

At two-loop level the situation is more involved because the H-diagram 
predicts~\cite{ACV90} the $D=4$ absorptive part
\begin{align}
  \Im\delta_2 = \Im a^{(H)} &= \log s \;\frac{Gs}{\hbar}(\pi R)^2
  \int\dif^2 x\;|h(b,x)|^2 \nonumber \\
  &= \frac{R^2}{\pi b^2}\left(\delta_0
    +\text{finite part}\right) \log s \;, \label{imdelta2} \\
    h(b,x) & \equiv \frac{x-x^*}{2\pi^2 b x^* (x-b)} \;,
\end{align}
where the field $h$ was introduced in~\cite{ACV07} and, in parallel with $h_s$,
is related to the metric coefficient $h_{zz^*}$ ($h_{z^* z^*}$ for $h_s$) of the
ACV metric~\cite{CCCV15}. Since $|h(x)|=\ord{1/|x|}$, the
result~\eqref{imdelta2} carries the logarithmic IR divergence parametrized by
$\delta_0$.
Furthermore, $2\delta_0\delta_1$ turns out to be of the same order as $\Im a^H$
by building up a total $\Im a^{(2)}$ in eq.~\eqref{ima2} which is 4 times larger
than $\Im a^{(H)}$.

That divergence is actually to be expected in the imaginary part, in order to 
compensate a similar divergence of virtual corrections, so as to yield a finite 
total emission probability. The trouble would be if the divergence of
$\Im a^{(H)}=\Im\delta_2$ were transferred to $\Re\delta_2$, because it would 
mean an IR singularity of a measurable quantity which is incurable, due to its 
multiplicative $b$-dependence.

Fortunately ACV were able to show that the IR divergence cancels out in 
$\Re\delta_2$, which is finite, thus leading to a no-renormalization argument 
for the infinite Coulomb phase at order $G^3 s^2$. In fact, by the same 
analyticity and asymptotic behaviour arguments used before, they found a unique 
solution to eqs.~\eqref{ima2} and \eqref{imdelta2} for $a^{(2)}$, given by the 
superposition of two analytic structures
\begin{equation}\label{a2v}
 a^{(2)} = \left(1+2\frac{\hbar^2 \nabla_{\bt}^2}{s}\right)
  \left(-\frac23\delta_0^3\right)
  +2\left(1+\frac{2\ui}{\pi}\log s\right)
  \left(\frac{\hbar^2 \nabla_{\bt}^2}{s}\Big(\frac23\delta_0^3\Big)
  +\frac{G^3 s^2}{\hbar b^2}\right) \;.
\end{equation}
Here the first term contains the leading iteration of the 2-body eikonal and
definite subleading contributions, while the second term contains also
the finite part of the H-diagram contribution, computed in~\cite{ACV90} in
dimensional regularization.  By working out the $\nabla^2$ terms, we can check
that the IR singular $\Im a^{(2)}$ is consistent with eq.~\eqref{ima2}, while
the divergence of the real part cancels out between the two terms. In
conclusion, we do not need any new divergent Coulomb phase at order $G^3 s^2$.
The outcome of the calculation is just the finite result%
\footnote{This relatively simple derivation, basically a recollection
of~\cite{ACV90}, can be seen as a shortcut resting on some plausible
analyticity assumptions and should not be taken as a substitute for a full
explicit --- and technically challenging --- calculation that we leave to
further work.}
\begin{equation}\label{redelta2}
  \Re\delta_2 = \frac{Gs}{\hbar}\frac{R^2}{2b^2} \;,
\end{equation}
which provides the first correction to both the eikonal and the Einstein
deflection angle at relative order $R^2/b^2$. In the Breit frame for scattering
ACV found the deflection
\begin{equation}\label{defl}
  \sin\frac12 \Theta_s(b) = \frac{R}{b}\left(1+\frac{R^2}{b^2}\right) \;.
\end{equation}

\section{Infrared logs in radiation and eikonal resummation\label{s:ilr}}

So far, following~\cite{CCCV15,GrVe14} we have constructed a graviton radiation
amplitude unifying the fragmentation and central emission regions in our eikonal
approach.  We have shown that the effect of the large-distance gravitational
interaction cancels out at the level of the (infinite) scattering phase. Here we
investigate the same question at the level of gravitational radiation.

Indeed, we meet immediately a possible problem at the single-graviton exchange
level. The amplitude (say, for helicity $\lambda=-2$) is directly related to the
field $h_s$ of eq.~\eqref{hsxxs} by a Fourier transform:
\begin{equation}\label{M1}
  \tfa^{(1)}_{\lambda=-2}(\bt,\qt) = \sqrt{\ag}\frac{R}{2}\frac{q^*}{q}
  \int\dif^2\xt\; \esp{\ui\qt\cdot\xt/\hbar} h_s(b,x) \;.
\end{equation}
Here the integral is linearly IR divergent by power counting, due to the
large-$\xt$ behaviour $\sim1/|\xt|$ of $h_s$ (and $h$). Nevertheless, the
Fourier Transform can be done thanks to the oscillating factor
$\esp{\ui\xt\cdot\om\tht}$ and yields the expression
\begin{align}
  \tfa^{(1)} &= \esp{\frac{\ui}{2}\qt\cdot\bt} \sqrt{\ag} \frac{R}{\pi}\;
  \Re\left[\esp{-\frac{\ui}{2}\qt\cdot\bt}\frac{\ui}{\pi}\left(
      \frac1{q^* b}-\int_0^\infty\frac{\dif t\;\esp{-t}}{q^* b-2\ui t}
    \right)\right] \nonumber \\
  &\stackrel{bq\to0}{\simeq}\sqrt{\ag}\frac{R}{\pi}\left[-\frac{\sin\phi_\qt}{|qb|}
    +\frac12\log\frac{2}{|qb|} +\text{const in }qb\right] \qquad
  (|\theta|\gg\theta_m) \label{M1bq}
\end{align}

We note that the expected soft behaviour $\sim1/\om$ is accompanied by a
logarithmic one, probably related to the proposal in~\cite{SaSe18} and that both
involve the variable $b\qt$ by showing a strong $\tht$-dependence, which is not
square-integrable at $\qt=\om\tht\to0$, and --- as it stands --- is not usable
for physical spectra.

In other words, here we stress the point that the single-exchange amplitude is
very sensitive to the IR in the span $0<|\xt|<\hbar/|\qt|$ and shows a spurious
singularity at $\qt=0$ due to large distances, despite the absence of collinear
singularities in the matrix element%
\footnote{This feature can be ascribed to the fact that the single-exchange
amplitude in $\bt$-space does not know anything about the angular scale
$\Theta_s = 2 R/b$ and is instead dominated by the very small-angle region
$\theta_m \sim \frac{\hbar}{Eb} \ll \Theta_s$.}.
But the way out this potential problem is just the correct use of the
single-exchange amplitude as an intermediate result, in order to calculate the
complete one. In fact, we know from start that we have to sum over all possible
exchanges in order to be able to reach physical deflection angles of order
$\Theta_E=2R/b\gg\theta_m=\hbar/(bE)$. Such resummation is possible because of
high-energy factorization~\cite{CCCV15} at fixed impact parameter $\bt$, and
takes into account the fact that the incidence angles of the various
contributions are rotated, so as to cover, eventually, the larger angular range
$\theta_m \ll|\tht|\sim 2R/b < 1$ they are required to describe. By working out
that procedure it was found~\cite{CCCV15} that the two contributions in
eq.~\eqref{svilamp} exponentiate independently by yielding the result
\begin{align}
  \tfa_\lambda &= \esp{2\ui\delta(b,E)} \; \ampRid_\lambda(\bt,E,\qt,\om)  \nonumber \\
  2\ui\om\ampRid_\lambda &= \frac{\sqrt{\ag}}{\pi}\esp{\ui \lambda \phi_\tht}\int
  \frac{\dif^2 \zeta}{2\pi |\zeta|^2}\esp{-\ui \lambda \phi_\zeta}\;\esp{\ui\zetat\cdot\tht} \left(
    \esp{-2\ui\om R\log\left|\hat{b}-\frac{\zetat}{\om b}\right|}
    -\esp{-\ui\zetat\cdot\Tht_s} \right)  \nonumber \\
  &= \frac{\sqrt{\ag}}{\pi}\esp{\ui \lambda \phi_\tht}\int
  \frac{\dif^2 \zeta}{2\pi |\zeta|^2}\esp{-\ui \lambda \phi_\zeta}\;\esp{\ui\zetat\cdot(\tht-\Tht_s)}
  \left(\esp{-2\ui\om R\Phic\big(\frac{\zetat}{b\om}\big)} - 1 \right)
 \;,\label{2iomb}
\end{align}
in terms of the rescaled variable $\zetat\equiv b\om\zt = \om\xt$.  This is in
complete agreement with the result of the fully classical calculation
of~\cite{GrVe14}.

We note that, because of~\eqref{2iomb}, resummation involves the phase factor
$\esp{-2\ui\om R\Phic}$ to keep coherence on the $\xt$-space involved.  In
practice that means that we should require, qualitatively, that
$1\leq|\Phic(\xt/b)|\leq 1/(2\om R)$ for coherence to be reached, thus reducing
the IR sensitivity span to $b<|\xt|<1/(\om\Theta_E)$. In other words, the
$|\tht|$-dependence is regularized around $\Theta_E=2R/b$, way before reaching
the IR singularity peak. As a consequence, our resummed amplitudes are finite in
the small-$\om$ region and well-behaved on the whole physical phase space.

Finally, we resum the independent emissions of many gravitons whose amplitudes
are factorized in terms of the emission factor $\ampRid$ of eq.~\eqref{2iomb}.
The $S$-matrix operator acting on the Fock space of gravitons is then obtained
by including virtual corrections which are incorporated by exponentiating both
creation ($a^\dagger_\lambda(\vq)$) and destruction ($a_\lambda(\vq)$) operators
of definite helicity $\lambda$ as follows
\begin{equation}\label{Shat}
  \hat{S}=\esp{2\ui\delta(\bt,E)}
   \exp\left\{\int\frac{\dif^3 q}{\sqrt{2\omega}}\;
    2\ui\left[\sum_\lambda \ampRid_\lambda(\bt,\vq) a^\dagger_\lambda(\vq)
      +\text{h.c.}\right] \right\} \;.
\end{equation}
Such a simple coherent state assumes negligible correlations among the emitted
gravitons, an assumption which is certainly justified by the factorization
theorems~\cite{We65} of multiple soft graviton emission. We believe this to be
still a good approximation in the region $\om R <1$ discussed in this work.
Such an $S$-matrix is unitary because of the hermitian operator appearing in the
exponent. It also conserves energy as long as we limit ourselves to processes in
which the total energy carried by the emitted gravitons is much smaller than
$\sqrt{s}$.

Given \eqref{Shat} it is straightforward to compute the energy carried by the
gravitons as a function of $\om$, $\tht$ and $\lambda$, in terms of the
expectation value of the corresponding operator
$\hbar \om a^\dagger_\lambda(\vq) a_\lambda(\vq)$. Using standard properties of
coherent states this is just given by
\begin{equation}\label{angspec}
  \frac{\dif E^\GW}{\dif\omega\, d^2 \tht} = 2 \hbar \sum_{\lambda=\pm 2}
  |\om\ampRid_\lambda(\bt,\vq)|^2 \;,
\end{equation}
which is directly related to the spectrum in the small-angle kinematics
\eqref{mompar} and has a smooth classical limit since $\ag$ is
$\ord{\hbar^{-1}}$. The explicit calculation will be carried out in
sec.~\ref{s:sob}.

\section{Small-\texorpdfstring{$\bs\om$}{w} behaviour of the radiation amplitude
\label{s:sob}}

In this section we will study the gravitational radiation spectrum
$ \frac{\dif E^\GW}{\dif\omega}$ in the small-$\omega$ region, here defined by
$\om R < (\ll) 1$.  Since, throughout this paper, we work at leading order in
the scattering angle $\Theta_s=2R/b \ll1$, this region is actually divided in
two subregions: $\om b < (\ll) 1$ and $\om b > (\gg) 1$.  In the complementary
regions $\om R=\ord{1}$ and $\om R > (\gg) 1$, analyzed in detail
in~\cite{GrVe14,CCV15,CCCV15}, decoherence effects --- related to the
exponentiation of $2\om R\Phic$ in eq.~\eqref{2iomb} --- suppress the
integration region%
\footnote{ Since we shall not use anymore complex notation for transverse
  vectors, from now on we denote the modulus of a transverse vector with the
  corresponding non-boldface symbol, e.g., $\theta\equiv|\tht|$.}
$\theta > \Theta_s$ and creates a break in the spectrum
around ``Hawking's frequency'' $\om\sim R^{-1}$, with a $1/(\om R)$ tail. The
whole treatment then becomes unreliable above $\om R \sim \Theta_s^{-2}$. We
will have nothing more to say here about the $\om R > 1$ regime.

On the other hand, in the small-$\om R$ region, there is a clear distinction
between the two above-mentioned (sub)regimes $\om b < (\ll) 1$ and $\om b >
(\gg) 1$.  Before turning to their quantitative study let us anticipate some
qualitative aspects of each.
\begin{itemize}
\item For $\om b < (\ll) 1$ we find corrections to the ZFL in the form of an
expansion in powers of $\om R$ which get enhanced by logarithms of $\om b$.
Even if small, these corrections (not considered in the earlier treatments
of~\cite{GrVe14,CCV15,CCCV15}) are obviously important for determining whether
the spectrum is (or  is not) maximal at the ZFL. Furthermore,  since the ZFL
itself is of $\ord{\Theta_s^2}$,  the $\ord{\om R \log(\om b)}$ and
$\ord{\om^2 R^2 \log^2(\om b)}$ leading corrections turn out to be of relative
order $\ord{\om b \log(\om b)}$ and  $\ord{\om^2 b^2 \log^2(\om b)}$,
respectively. The first one, while representing an interesting memory effect on
the wave form and a contribution to the polarized flux,  does not contribute to
the unpolarized and/or azimuthally-averaged flux. The second represents
instead the leading contribution to the unpolarized and/or angle-integrated
flux. Its positivity implies necessarily a maximum of the spectrum away from
$\om =0$. Finally, we  will be able to resum all the leading logs in terms of
an IR-finite Coulomb phase.
\item For $\om b > 1$ the above-mentioned logarithmic enhancements disappear
and, instead, a cutoff intervenes at $\theta \sim (\om b)^{-1} $. As a result,
the maximum of the spectrum is reached at $\om b = \ord{1}$: numerically, it is
found to stay,  independently of $\Theta_s$, around $\om b \sim 0.5$. For
$\om b \gg 1 \gg \om R$ a previously studied  regime  settles in, in which the
spectrum decreases like $\log(1/\om R)$~\cite{GrVe14,CCV15,CCCV15}.
\end{itemize}

After recasting eq.~\eqref{2iomb} in a more convenient form, we shall recover,
in sec.~\ref{s:la}, the leading-$\om$ contributions in the region $0 <\om R <1$,
while in sec.~\ref{s:sla} we compute the new sub-leading corrections and the
emergence of a peak in the spectrum at $\om b \sim 0.5$.

We start by defining
\begin{equation}\label{A}
  A^{(\lambda)} \equiv \frac{2 \pi \ui\om\ampRid_{\lambda} }{ \sqrt{\ag}}
  = A^{(\lambda)}_L  + A^{(\lambda)}_{NL}
\end{equation}
where, using eq.~\eqref{2iomb}, we can identify:
\begin{equation}\label{AL}
  A^{(\lambda)}_L = \int_{0}^{\infty}\frac{\dif r}{r}
   \int_0^{2\pi}\frac{\dif\phi_x}{2\pi}\;
  \esp{\ui\lambda (\phi_{\tht} -\phi_x)}
  \left( \esp{\ui\om\xt\cdot\tht} - \esp{\ui\om\xt\cdot(\tht-\Tht_s)} \right)
\end{equation}
and
\begin{equation}\label{ANL}
  A^{(\lambda)} _{NL} = \int_{0}^{\infty} \frac{\dif r}{r}
   \int_0^{2\pi} \frac{\dif\phi_x}{2\pi}\;
  \esp{\ui \lambda (\phi_{\tht} -\phi_x)}  \esp{\ui \om \xt\cdot\tht}
  \left( \esp{-\ui \om R \log( \hat{\bt} - \frac{\xt}{b})^2} -1 \right) \;.
\end{equation}
In order to evaluate $A_{NL}$, we will use the expansion:
\begin{equation}\label{explog}
  \log( \hat{\bt} - \frac{\xt}{b})^2 = \log(1 + \frac{r^2}{b^2})
  + \log(1 - 2 \frac{\bt\cdot\xt}{b^2 +r^2})
  = \log(1 + \frac{r^2}{b^2}) - 2  \frac{\bt\cdot\xt}{b^2 +r^2} + \dots \, ,
\end{equation} 
which is valid both at large and at small $r/b$.
By replacing~\eqref{explog} into~\eqref{ANL}, we rewrite
\begin{equation}\label{ANLi2}
  A^{(\lambda)} _{NL} \simeq \int_{0}^{\infty} \frac{\dif r}{r}
   \int_0^{2\pi} \frac{\dif\phi_x}{2\pi}\;\esp{\ui \lambda (\phi_{\tht}-\phi_x)}
   \left[\esp{\ui\om\xt\cdot(\tht-\Thr)}\esp{-\ui\om R\log(1+\frac{r^2}{b^2})}
     - \esp{\ui\om\xt\cdot\tht}\right] \;,
\end{equation}
where we have introduced what we call the rescattering deflection angle
$\Thr\equiv\Tht_s / (1+r^2/b^2)$ which, together with the eikonal phase
$\om R\log(1+r^2/b^2)$, describes the rescattering evolution of the emitted
graviton.

\begin{figure}[ht]
  \centering
  \includegraphics[width=0.3\linewidth]{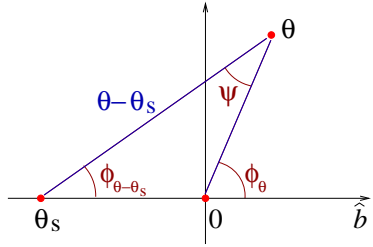}
  \caption{\it Picture of the polar and azimuthal angles in the transverse
    plane. $\Tht_s$ and $\tht$ correspond respectively to the projections of the
    unit-vectors $\hat{p}_1{}'$ and $\hat{q}$ on the $\bk{x,y}$-plane of
    fig.~\ref{f:3DimpactParam}. In this configuration, all azimuthal angles
    $\phi_j$ and $\psi$ are positive.}
  \label{f:azimutali}
\end{figure}

We then split the $r$-integration into two regions: $r < (\ll)b$, and 
$r>(\gg)b$. In the small-$r$ region, $\Thr\simeq\Tht_s$, the
$\Tht_s$-dependence cancels out between $A_L$ and $A_{NL}$ and can be eliminated
in their sum. Performing now the azimuthal integrations in terms of  Bessel
functions, we obtain
\begin{equation}\label{AL1}
  A_L^{(\lambda)} = \int_{b}^{\infty} \frac{\dif r}{r} \left[\esp{ \ui \lambda \psi}
    J_2(\om r |\tht - \Tht_s|) - J_2(\om r \theta) \right] \;,
  \qquad \psi \equiv (\phi_{\tht} - \phi_{\tht-\Tht_s}) \;,
\end{equation}
where $\psi$ is the azimuthal-angle transfer in scattering
(see fig.~\ref{f:azimutali}), and
\begin{equation}\label{ANL1}
  A_{NL}^{(\lambda)} \simeq \int_{b}^{\infty} \frac{\dif r}{r}\left[J_2(\om r\theta)
  - \esp{\ui\lambda\psir} \esp{-\ui\om R\log(1+\frac{r^2}{b^2})}
    J_2(\om r |\tht - \Thr|) \right] \;,
  \qquad \psir \equiv (\phi_{\tht} - \phi_{\tht-\Thr}) \;,
\end{equation}
where $\psir$ is the analogue azimuthal transfer in rescattering. Furthermore,
the $r$-integration is now limited to the large-$r$ region.

Since $\Thr\ll 1$ in the large-$r$ region, we neglect it in the argument of the
Bessel function in eq.~\eqref{ANL1}, to get the simplified form
\begin{align}
  A_{NL}^{(\lambda)} &\simeq \int_{b}^{\infty} \frac{\dif r}{r}\; J_2(\om r\theta)
  \left[ \left(1 - \esp{-\ui\om R\log(1+\frac{r^2}{b^2})} \right)
    + \esp{-\ui\om R\log(1+\frac{r^2}{b^2})}
    (1 - \esp{\ui\lambda\psir})\right]\nonumber\\
& \equiv A_{NL,C} +\delta A_{NL}^{(\lambda)} \;, \label{ANL2}
\end{align}
where the eikonal-phase contribution $A_{NL,C}$ is the main one to be discussed
below, while the rescattering phase can be further expanded to first order in
$\Thr$:
\begin{equation}\label{eikphase}
  1-\esp{\ui\lambda\psir} \simeq -\ui\lambda\psir
  = -\ui\lambda\frac{\Theta_s \sin\phi_\tht}{\theta(1+\frac{r^2}{b^2})}
\end{equation}
and is correspondingly small. By replacing that value into~\eqref{ANL2} we
obtain
\begin{align}\label{dANL}
  \delta A_{NL}^{(\lambda)} &\simeq -\ui\lambda
  \frac{\Theta_s \sin\phi_\tht}{\theta}
  \int_b^\infty\frac{\dif r}{r}\;\frac{J_2(\om r\theta)}{1+\frac{r^2}{b^2}}
  \esp{-\ui\om R\log(1+\frac{r^2}{b^2})} \nonumber \\
  &\stackrel{b\om\ll 1}{=}
  -\ui\lambda \frac{\Theta_s \sin\phi_\tht}{\theta} \left[
    \frac{(b\om\theta)^2}{4}\log\frac{C}{b\om\theta}
    + \om R\text{-corrections} \right] \;,
\end{align}
where the latter estimate comes from the small-$x$ Bessel expansion
$J_2(x)\simeq x^2/4$ and $C$ parametrizes the upper limit of that regime.

We thus see that there is a logarithmic enhancement of the nominal
$(b\om\theta)^2$ behaviour of $\delta A_{NL}^{(\lambda)}$, but is not the
maximal one. For that reason, in the following we shall mostly focus on the term
$A_{NL,C}$ of~\eqref{ANL2}, which will be shown to contain leading-log
contributions and to be related to the Coulomb phase of rescattering.

By then leaving $\delta A_{NL}^{(\lambda)}$ aside for the time being, and with
the approximation $\log(1+\frac{r^2}{b^2})\simeq\log(\frac{r^2}{b^2})$ in the
exponents, we can write
\begin{align}\label{finalA}
  A^{(\lambda)} -\delta A_{NL}^{(\lambda)} &= \int_b^\infty \frac{\dif r}{r} \;
  \left[\esp{ \ui\lambda\psi}
    J_2(\om r |\tht - \Tht_s|) - J_2(\om r \theta) \right] \nonumber \\
  &\quad + \int_b^\infty \frac{\dif r}{r}
  \left( 1 - \esp{- \ui \om R \log(\frac{r^2}{b^2})}\right)
  J_2(\om r \theta) \nonumber \\
  &= \left[\esp{ \ui\lambda\psi}
    \frac{J_1(\om b |\tht - \Tht_s|)}{\om b |\tht - \Tht_s|} -
    \frac{J_1(\om b \theta)}{\om b \theta} \right]
  + 2 i \om R \int_{\om b \theta}^{\infty} 
  \frac{\dif\rho}{\rho^2} \;J_1(\rho)
  \left(\frac{\rho}{b\om\theta}\right)^{-2 \ui\om R}\;,
\end{align}
where we have used a well-known Bessel integral and the last term (carrying an
explicit $\om R$ factor) is obtained through an integration by parts.  In the
following two subsections we will stick, for simplicity, to this simpler
analytic approximation which is sufficient to discuss the qualitative feature of
the spectrum. However, in sec.~\ref{s:num}, we will compare numerical results
with the better approximation given in eqs.~\eqref{AL},\eqref{ANLi2}.

Note that the amplitudes for $\lambda = \pm 2$ are {\em not} each other's
complex conjugates.  Equation~\eqref{finalA} is a convenient starting point for
analyzing various limits. In particular, the subleading corrections enhanced at
leading logarithmic level come from the last term.

\subsection{The leading amplitude for
  \texorpdfstring{$\bs{\om R < 1}$}{w R < 1} \label{s:la}}

Inspection of the small-$\om$ behaviour of the last term in eq.~\eqref{finalA}
shows that it vanishes in the $\om \rightarrow 0$ limit.  Limiting ourselves to
the first two terms we note first that the $J_1$ terms are leading and close to
$1/2$ for small values of the argument $\ord{bq}<1$. That is, for
$1>\theta>\Theta_s$ eq.~\eqref{finalA} becomes
\begin{equation}\label{leadAmp}
  A_L^{(\lambda)} \simeq \frac{J_1(\om b \theta)}{\om b \theta}
  \left(\esp{ \ui\lambda\psi} -1 \right)
\end{equation}
and yields
\begin{align}
  \frac{\dif E_{\lambda}^\GW}{\dif\om\dif\Omega} =
  2 \hbar |\om\ampRid_{\lambda}|^2
  &\simeq \frac{Gs}{8\pi^2} \left|\frac{2J_1(bq)}{bq}\right|^2
  \left|\esp{\ui \lambda \psi}-1\right|^2
  \nonumber \\
  &= \frac{Gs}{2\pi^2}\left|\frac{2J_1(bq)}{bq}\right|^2
   \, \frac{\sin^2\phi_\tht}{|\tht-\Tht_s|^2}\,\Theta_s^2 \;.\label{spect}
\end{align}
where we have used the trigonometric relation (see fig.~\ref{f:azimutali}):
$\sin \psi = \frac{\Theta_s}{|\tht - \Tht_s|} \sin \phi_{\theta}$.

On the other hand, $bq>1$ is allowed by phase space if $b\om>1$, and in that
case the $J_1$ factors suppress the amplitude, consistently with previous
estimates~\cite{CCCV15} of the large $bq$ behaviour.  By integrating over the
angular phase space%
\footnote{Because of the forward-backward symmetry of the process, graviton
  radiation in the backward hemisphere occurs at the same rate. In practice, in
  the small-angle kinematics,
  $\int_{\sphere_2}\dif\Omega=2\int_0^1\dif\theta\;\theta\int_0^{2\pi}\dif\phi$.}
we then find the $\lambda$-independent result:
\begin{align}
  \frac{\dif E_{\lambda}^\GW}{\dif\om} &\simeq \frac{Gs}{\pi}\int_0^1
  \frac{\dif\theta}{\theta}\;
  \left|\frac{2J_1(b\om\theta)}{b\om\theta}\right|^2
  \min\{\theta^2,\Theta_s^2\} \nonumber \\
  &\simeq \frac{Gs}{\pi}\int_0^{\min\{1,1/b\om\}}
    \frac{\dif\theta}{\theta}\; \min\{\theta^2,\Theta_s^2\} \nonumber \\
    &= \frac{Gs}{\pi}\Theta_s^2 \left[
      \log\min\big(\frac1{\Theta_s},\frac1{2\om R}\big)+\text{const}\right] \;.
    \label{specb}
\end{align}
Note that the spectrum takes the ZFL form for $\om<b^{-1}$ but differs by the
phase-space condition $\theta<(b\om)^{-1}$ (or $bq<1$) for $\om>b^{-1}$,
as required by the large-log assumption. As a consequence, the full frequency
spectrum has a $\log(1/\om R)$ dependence of the form:
\begin{equation}\label{specc}
  \frac{\dif E^\GW}{\dif\om} = \frac{Gs}{\pi}\Theta_s^2
  \left[ \text{const} + 2\log\big(\frac{1}{2\om R}\big)\right] \;,
\end{equation}
which saturates at $\om=b^{-1}$, reaching the ZFL value. 

\subsection{Subleading corrections and IR-sensitive logs \label{s:sla}}

Enhanced subleading corrections come entirely from the last term in
eq.~\eqref{finalA}. As a matter of fact that term is known exactly in terms of
an hypergeometric function:
\begin{align}
  & 2 \ui \om R \int_{\om b |\theta|}^{\infty} \frac{\dif\rho}{\rho^2} \;
  J_1(\rho) \left(\frac{\rho}{b \om \theta}\right)^{-2 \ui \om R}
    \equiv \Delta_C(\om R, \om b \theta) \nonumber \\
  &= \frac12  \hyp{-\ui\om R}{2}{1-\ui\om R}{ -\frac{\om^2 b^2 \theta^2}{4}}
    - \frac12 \frac{\Gamma(1-\ui\om R)}{\Gamma(1+\ui\om R)}
    \frac{(\frac{b\om\theta}{2})^{2 \ui \om R}}{1+ \ui \om R}
 \label{lasterm}
\end{align}

We may now collect all terms in $A^{(\lambda)}$,
 \begin{equation}\label{AA}
   A^{(\lambda)}  = \left[\esp{\ui\lambda\psi}
     \frac{J_1(\om b |\tht - \Tht_s|)}{\om b |\tht - \Tht_s|}
     -\frac{J_1(\om b \theta)}{\om b \theta} \right]  
   + \Delta_C(\om R, \om b \theta)
\end{equation}
and note that the two definite-helicity amplitudes differ just by an imaginary
term proportional to $\sin 2 \psi$.  On the other hand, if we consider the more
conventional linear polarizations:
\begin{equation}\label{xand+}
  A^{(\times)} = A_{LT} = \frac{i}{\sqrt{2}} (A^{(2)} - A^{(-2)}) \;;\qquad
  A ^{(+)} = A_{TT} = \frac{1}{\sqrt{2}} (A^{(2)} + A^{(-2)})
\end{equation}
we see that the $\lambda$-dependent term in eq.~\eqref{AA} only contributes to
$A^{(\times)}$.

\subsubsection{Small \texorpdfstring{$\bs{\om b}$}{wb} regime}
Before moving on to a discussion of the spectrum at generic values of $\om b$,
let us consider the small $\om b$ limit. In that limit we saw that the single
emission amplitude contains a divergent $\log(1/bq)$ at subleading (in $\omega$)
level. However, the resummed amplitude is finite, in fact the large logarithms
appear in $\Delta_C$ in the resummed exponential form
\begin{equation}\label{lastermexp}
  \Delta_C(\om R,bq) \sim \frac12 -\frac12 \esp{-2\ui\om R\log(2/bq)} \;,
\end{equation}
yielding an oscillatory function. By adding the leading term, the small $\om b$
limit of the amplitude reads
\begin{equation}\label{2ioMsob}
  2\ui\om\ampRid_\lambda \simeq \frac{\sqrt{\ag}}{\pi}\frac12\left[
    \esp{\ui\lambda\psi}-\esp{-2 \ui \om R\Lbq} \right] \;, \qquad
  \Lbq \equiv \log(2/b\om\theta)
\end{equation}
or, equivalently,
\begin{subequations}\label{A+-}
\begin{align}
  A^{(2)} &= \ui[ \esp{\ui\psi} \sin\psi
    + \esp{-\ui\om R\Lbq}\sin(\om R\Lbq)] \;, \\
  A^{(-2)} &= \ui[ -\esp{-\ui\psi}\sin\psi
    + \esp{-\ui\om R\Lbq}\sin(\om R\Lbq)] \;.
\end{align}
\end{subequations}
For the linear polarizations we find
\begin{subequations}\label{ALT}
\begin{align}
  A^{(+)} &= -\sqrt{2}\sin^2\psi + \ui\sqrt{2}\sin(\om R\Lbq)
  \esp{-\ui\om R\Lbq} =\sqrt2 [\Re A_L + A_{NL}]\;, \\
  A^{(\times)} &= -\sqrt{2}\sin\psi\cos\psi
  = \sqrt2\, \Im A_{L}^{(-2)} \;.
\end{align}
\end{subequations}
As a consequence, the interference patterns at fixed helicity are of the form
\begin{equation}\label{Asq}
  |A^{(\pm2)}|^2 = \sin^2\psi + \sin^2(\om R\Lbq) \pm 2\sin\psi
  \sin(\om R\Lbq) \cos(\psi\pm\om R\Lbq) \;.
\end{equation}
We can see that interference starts at leading order $\sim\om R\Lbq$, has
opposite sign for the two helicities, and cancels out after azimuthal
integration in $\phi_\tht$. On the other hand, if only the total (unpolarized)
energy flux is measured, we get
\begin{equation}\label{enflux}
  |A^{(2)}|^2 + |A^{(-2)}|^2 = |A^{(+)}|^2 + |A^{(\times)}|^2
  = 2 \left[ \sin^2\psi \cos^2(\om R\Lbq)
  + \sin^2(\om R\Lbq) \cos^2\psi\right] \;,
\end{equation}
showing no first-order interference.

The same conclusions can be drawn by recalling that the two helicity
amplitudes~\eqref{AA} differ just by a term proportional to $\sin 2 \psi$. By
taking into account the relations:
\begin{align}
  \sin \psi &= \frac{\Theta_s}{|\tht - \Tht_s|} \sin \phi_{\tht}  \;; \quad
  \cos \psi = \frac{\theta - \Theta_s \cos \phi_{\tht}}{|\tht - \Tht_s|}
  \nonumber \\
  & |\tht-\Tht_s|=\sqrt{\Theta_s^2+\theta^2-2\theta\Theta_s\cos\phi_{\tht}}\,,
 \label{angles}
\end{align}
we can check that the azimuthal average of $\sin 2 \psi$ vanishes.  Since the
other terms in $A^{(\lambda)}$ do not depend on $\phi_{\tht}$, we conclude that
the azimuthal average of the energy flux is the same for the two
helicities. Also, in the total flux there is no term linear in $\sin 2 \psi$
that survives.

Furthermore, we notice that a similar resummation can be performed on the
subleading log amplitude $\delta A_{NL}$ by using~\eqref{dANL} at higher orders
in $\om R$, to yield
\begin{equation}\label{dANLris}
  \delta A_{NL}^{(\lambda)} = -\ui\sgn(\lambda) b\om\theta \sin\phi_\tht
  \sin(\om R\Lbq) \esp{-\ui\om R\Lbq} \;.
\end{equation}
Since this contribution has opposite values for the two helicities, it doesn't
affect the $A^{(+)}$ polarization and contributes only to $A^{(\times)}$, which
becomes
\begin{equation}\label{AxNL}
  A^{(\times)} = -\sqrt{2}\sin\psi\cos\psi + \sqrt{2} b\om\theta \sin\phi_\tht
  \sin(\om R\Lbq) \esp{-\ui\om R\Lbq} \;.
\end{equation}
The corresponding change to the unpolarized the energy flux~\eqref{enflux} is
given by the square of the second term in eq.~\eqref{AxNL}, which we neglect
being of order $(\om R)^4$, and by the interference of the two terms in the same
equation, which is of order $(\om R)^2$, like the last term in
eq.~\eqref{enflux}, and reads
\begin{equation}\label{dAsq}
  \delta|A^{(\times)}|^2
  = - b\om\theta\sin\phi_\theta \sin(2\psi) \sin(2\om R\Lbq) + \ord{\om R}^4 \;.
\end{equation}
By performing the azimuthal average of eqs.~\eqref{enflux} and \eqref{dAsq}
using the elementary integral:
\begin{equation}\label{elint}
  \int_0^{2\pi} \frac{\dif\phi}{2\pi}\;\sin^2\phi\frac{\Theta_s^2}{|\tht-\Tht_s|^2}
  = \frac12 \left( \Theta_{H} (\Theta_s-\theta)
  + \Theta_{H} (\theta -\Theta_s) \frac{\Theta_s^2}{\theta^2} \right),
\end{equation}
where $\Theta_{H}$ is the Heaviside step-function, we obtain
\begin{align}
  \frac{\dif E^\GW}{\dif\om\,\dif\theta\,\theta} &= 2\frac{Gs}{\pi}
  \left\{\Theta_H(\Theta_s-\theta)\left[1-\om R\sin(2\om R\Lbq)
    \frac{\theta^2}{\Theta_s^2}\right] \right. \label{polarflux}\\
  &+ \left. \Theta_H(\theta-\Theta_s)
  \left[2\sin^2(\om R\Lbq)+\cos(2\om R\Lbq)\frac{\Theta_s^2}{\theta^2}
    -\om R\sin(2\om R\Lbq)\left(2-\frac{\Theta_s^2}{\theta^2}
    \right) \right] \right\} \;.  \nonumber
\end{align}
The contribution of the NL correction~\eqref{dAsq} to the previous expression is
given by the last terms (with the $\om R$ factor) in square brackets: they
provide a negative definite correction to the energy flux stemming from
eq.~\eqref{enflux}.

In order to study the small $\Theta_s$, small $\om b$ limit, it is useful to
expand $\Delta_C$ as:
\begin{equation}\label{lastermexpansion}
  \Delta_C(\om R,\om b\theta) \simeq + \ui \om R \Lbq + \om^2 R^2 \Lbq^2 + \dots
\end{equation}
Only the first term of the expansion turns out to be relevant in this limit. In
order to show this let us collect the leading contributions to eq.~\eqref{AA}:%
\footnote{On the other hand, for $\om R \ll 1, \om b \theta \gg 1$ no large logs
  survive (they cancel between the two terms in~\eqref{lasterm}) and, instead,
  $|\Delta_C|$ effectively provides a cutoff at $ \om b \theta \sim 1$.}
\begin{equation}\label{Aexp}
  A ^{(\lambda)} \sim \left[- \sin^2(\psi) \left(1- \frac18 \om^2 b^2 \theta^2\right)
    + \om^2 R^2 \Lbq^2 \right]
    + \ui \left[\frac{ \sin(\lambda\psi)}{2}
    \left(1- \frac18 \om^2 b^2 \theta^2\right) + \om R \Lbq \right]
\end{equation}
Taking now the absolute square of~\eqref{Aexp}, and isolating contributions of
order $\Theta_s^2$, we see that the real part can be neglected. From the
imaginary part, the leading term in $\om$ comes from squaring the
$\sin(\lambda \psi)$ with a correction
$\ord{\sin\phi_\tht\,\om b\log(b\om\theta)}$ originating from its interference
(that cancels after azimuthal averaging) with the last term and, finally, a
correction $\ord{\om^2 b^2 \log^2 (b\om\theta)}$ coming from squaring that same
term.  Higher order terms in the expansion of~\eqref{lastermexp} only contribute
to higher orders in $\Theta_s$.
 
It is also clear that the leading contribution comes from the $\lambda$-odd
component of $A ^{(\lambda)}$, the last correction from a $\lambda$-even term,
while the interference term needs both. As a result, there is no such
interference term for the linear polarizations, while such a correction exists
(with opposite contributions) for the two circular polarizations (helicities),
but vanishes upon integration over the azimuthal angle. Furthermore, the leading
term appears only in the $(\times) = LT$ polarization, while the $\ord{\om^2}$
correction only contributes to the $(+)=TT$ flux.

Our $\ord{\om}$ results can be compared with the ones obtained in
\cite{LaSe18,SaSe18} through subleading corrections to the soft-graviton
theorems. In that work one has to introduce by hand a recipe for regularizing an
IR infinity. When this is done there is perfect agreement between the two
calculations,%
\footnote{B.~Sahoo and A.~Sen, private communication. One of us
(GV) would like to thank Ashoke Sen for several discussions about how the first
  subleading correction contributes to different polarizations.}%
which can be seen as a confirmation of their recipe and as a way to fix the
scale of the $\log \om$ corrections. On the other hand, to the best of our
knowledge, the $\ord{\om^2}$ corrections are calculated here for the first time.

\subsubsection{Generic \texorpdfstring{$\bs{\om b}$}{wb}}
Let us now go back to~\eqref{AA} and to the case of generic values of $\om b$
considering the total flux (summed over the two polarizations).
Using~\eqref{angspec} and \eqref{AA}, we can write:
\begin{align}
  &\frac{2\pi^2}{\alpha_G} \frac{\dif E^\GW}{\dif\om\dif\Omega} = 2\left(
  \cos(2\psi) \frac{J_1(\om b |\tht - \Tht_s|)}{\om b |\tht - \Tht_s|}
  - \frac{J_1(\om b \theta)}{\om b \theta}
  +\Re\Delta_C(\om R, \om b \theta) )\right)^2 \nonumber \\
  & + \sum_{\lambda} \left(\sin(\lambda \psi)
  \frac{J_1(\om b |\tht - \Tht_s|)}{\om b |\tht - \Tht_s|}
  + \Im\Delta_C(\om R, \om b \theta) \right)^2
  \label{diffspect}
\end{align}
with $\Delta_C(\om R, \om b \theta)$ defined in eq.~\eqref{lasterm}. Note that
the only $\lambda$-dependence comes from the interference term in the square of
the imaginary part.  Because of~\eqref{lastermexpansion} this term is already
there at order $\om b$ but, as already mentioned, it disappears after either
integration over $\phi$ or after summing over $\lambda$.  Performing the latter
operation we arrive at:
\begin{align}
  &\frac{\pi^2}{2\alpha_G}
  \frac{\dif E^\GW}{\dif\om\dif\theta\,\theta\,\dif\phi}
  = 4 \sin^2 \psi \cos^2 \psi \left(
  \frac{J_1(\om b |\tht - \Tht_s|)}{\om b |\tht - \Tht_s|}\right)^2
  + \left(\Im\Delta_C(\om R, \om b \theta)\right)^2  \nonumber  \\
  & + \left[-2 \sin^2 \psi
    \frac{J_1(\om b |\tht - \Tht_s|)}{\om b |\tht - \Tht_s|} +
    \left(\frac{J_1(\om b |\tht - \Tht_s|)}{\om b |\tht - \Tht_s|}
    -\frac{J_1(\om b \theta)}{\om b \theta} \right)
    +\Re\Delta_C(\om R, \om b \theta) \right]^2  \nonumber \\
  & = \left(\Im\Delta_C \right)^2 + \left(
  \frac{J_1(\om b |\tht - \Tht_s|)}{\om b |\tht - \Tht_s|}
  - \frac{J_1(\om b \theta)}{\om b \theta} + \Re\Delta_C \right)^2 \nonumber\\
  &+4\sin^2\psi \frac{J_1(\om b |\tht - \Tht_s|)}{\om b |\tht - \Tht_s|}
  \left( \frac{J_1(\om b \theta)}{\om b \theta} - \Re\Delta_C \right)
  \label{diffspectthetaphi}
\end{align}
Before integrating over $\phi$ let us make some approximations that are valid to
leading order in the deflection angle $\Theta_s$. Noting that $\Re \Delta_C$ is
of order $\Theta_s^2$ (see~\eqref{lastermexpansion}), we can neglect it
everywhere in the last expression since it is either squared or it multiplies
quantities that vanish as $\Theta_s \to 0$.  We can now perform the $\phi$
integration and obtain:
\begin{align}
  \frac{\pi}{4\alpha_G} \frac{\dif E^\GW}{\dif\om\dif\theta\,\theta}
  = \left(\Im\Delta_C \right)^2 + \int_0^{2\pi} \frac{\dif\phi}{2 \pi}
  \left( \frac{J_1(\om b |\tht - \Tht_s|)}{\om b |\tht - \Tht_s|}
  - \frac{J_1(\om b \theta)}{\om b \theta} \right)^2
  + 4 \frac{J_1(\om b \theta)}{\om b \theta} I
   \label{diffspecttheta}
\end{align}
where
\begin{equation}\label{I}
  I = \int_0^{2\pi} \frac{\dif\phi}{2 \pi} \sin^2 \psi
  \frac{J_1(\om b |\tht - \Tht_s|)}{\om b |\tht - \Tht_s|}
  = \int_0^{2\pi} \frac{\dif\phi}{2 \pi} \sin^2\phi
  \frac{\Theta_s^2}{|\tht-\Tht_s|^2}\,
  \frac{J_1(\om b |\tht - \Tht_s|)}{\om b |\tht - \Tht_s|} \;.
\end{equation}
This last integral can be estimated by noting that the $\phi$ dependence in
$(J_1(x)/x)$ can be neglected both for $\theta > \Theta_s$ and for $\theta <
\Theta_s$ (in this latter case since $x$ is small) and by then
using~\eqref{elint}.  Finally, the second term in~\eqref{diffspecttheta} can be
estimated at order $\Theta_s^2$ by expanding to first order $J_1(x)/x$ with the
result:
\begin{equation}\label{J/xexp}
  \int_0^{2\pi} \frac{\dif\phi}{2 \pi} \left(
  \frac{J_1(\om b |\tht - \Tht_s|)}{\om b |\tht - \Tht_s|}
  -  \frac{J_1(\om b \theta)}{\om b \theta} \right)^2
  \sim \frac12 \om^2 b^2 \Theta_s^2
  \left(\frac{J_2(x)}{x}\right)_{x= \om b \theta}^2
\end{equation} 

In conclusion we can write:
\begin{align}
  \frac{\pi}{4G s} \frac{\dif E^\GW}{\dif\om\dif\theta\,\theta}
  &\sim (\Im \Delta_C)^2 +\frac12 \om^2 b^2 \Theta_s^2
  \left(\frac{J_2(x)}{x}\right)^2 \nonumber \\
  &\quad +\frac12 \Theta_{H} (\Theta_s-\theta) + 2 \Theta_{H} (\theta -\Theta_s)
  \frac{\Theta_s^2}{\theta^2} \left(\frac{J_1(x)}{x} \right)^2  \;,
  \qquad x \equiv \om b \theta
  \label{diffspecttheta1}
\end{align}
which goes over to~\eqref{polarflux} in the small-$\om b$ limit.

Before proceeding further let us note again (see the above discussion of the
small $\om b$ case) that there is just one contribution that dominates
$\om$-dependence at small $\om b$. This is the term $(\Im \Delta_C)^2$ which,
is positive and, according to~\eqref{lastermexpansion}, of order
$\om^2 b^2 \log^2 (\om b \theta)$. It is thus already clear that the spectrum
cannot have its absolute maximum at $\om =0$.%
\footnote{We are making here the implicit assumption that the large-$\theta$
  region does not given logarithmically enhanced corrections.}

The above differential spectrum is supposedly accurate at $\theta \ll 1$ but
suffers, in general, from corrections of relative order $\theta$. Therefore, we
can only compute the absolute normalization of those contributions to the total
flux $\frac{\dif E^\GW}{\dif\om}$ which are dominated by the small-$\theta$
behaviour of eq.~\eqref{diffspecttheta1}. An example of this kind is the
$\log(\Theta_s)$-enhanced contribution to the ZFL given in eq.~\eqref{specb}.
Another example is the dominant term of order $\log(1/\om R)$ at
$b^{-1} \ll \om \ll R^{-1}$ (see again eq.~\eqref{specb}), since in this case
there is an effective cutoff in $\theta$ at $(\om b)^{-1} \ll 1$. By contrast,
the coefficient of the leading $\om$-dependent correction --- hence the position
of the maximum --- is {\em not} dominated by the (very)-small-$\theta$ region
and is therefore determined with some (possibly sizeable) uncertainty.

\subsection{Numerical results \label{s:num}}

In this subsection we present numerical results that can be obtained by direct
numerical integration of the full eikonal model~\eqref{2iomb} and compare them
with those based on numerically integrating the analytic approximations
discussed in sec.~\ref{s:sob}.  We will concentrate our attention, in
particular, on $\dif E^\GW/\dif\om$, the frequency-spectrum of gravitational
radiation integrated over solid angle (with the proviso mentioned at the end of
sec.~\ref{s:sla}) and summed over the two polarizations.

\begin{figure}[hpt]
  \centering
  \includegraphics[height=0.75\linewidth,angle=-90]{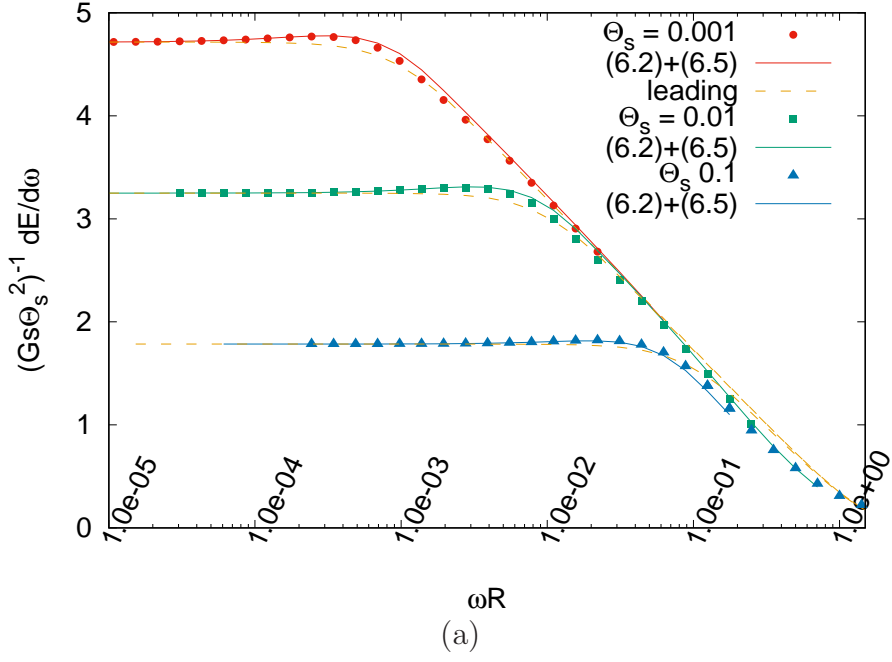}\\
  (a)\\
  \includegraphics[height=0.75\linewidth,angle=-90]{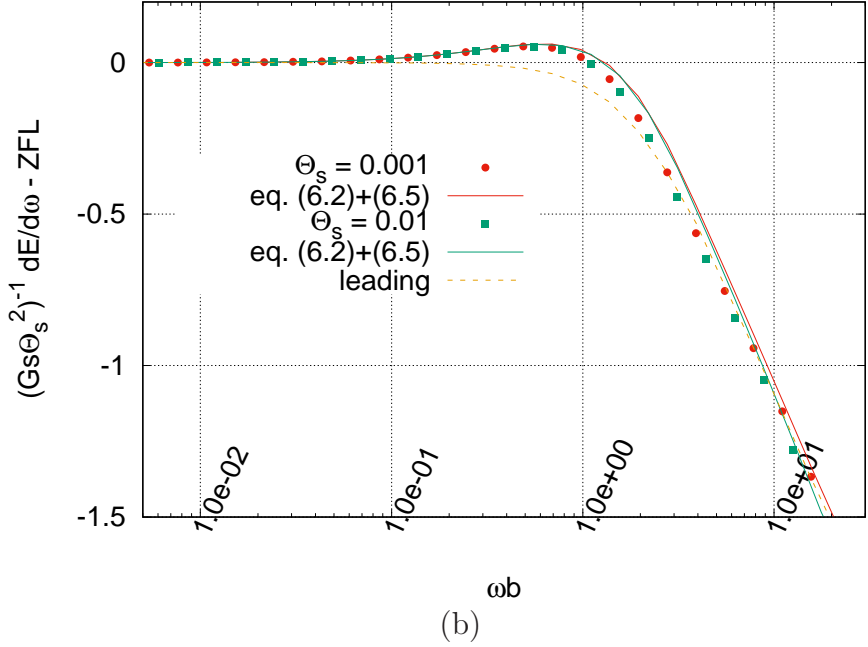}\\
  (b)
  \caption{\it {\rm(a)} The (reduced) graviton frequency spectrum against $\om R$ for
    three values of $\Theta_s$. Dots represent the full spectrum, while the
    solid lines represent the values obtained by using the analytic
    approximation~\eqref{AL}+\eqref{ANLi2} of the amplitude. The orange-dashed
    lines represent the leading approximation.  {\rm(b)} The (reduced) graviton
    frequency spectrum versus $\om b$ and with ZFL subtracted out, for two values of
    $\Theta_s$. The meaning of dots and lines is as in {\rm(a)} .}
  \label{f:spectrum}
\end{figure}

First of all we want to asses the validity of our approximations, which we use
to derive the main features of the radiation in the infrared region $\om R < 1$.
In the first plot (fig.~\ref{f:spectrum}a) we compare spectra%
\footnote{Actually, we plot a ``reduced'' spectrum with the kinematical factor
$Gs\Theta_s^2$ factored out.}
obtained with three values of the scattering angle
$\Theta_s = 10^{-1}, 10^{-2}, 10^{-3}$.  The points represent the spectra
calculated by numerical integration of the full amplitude~\eqref{2iomb} while
the solid lines are obtained by using the NL approximate
amplitude~\eqref{AL}+\eqref{ANLi2}. The orange-dashed lines correspond to the
leading approximation~\eqref{AL}-\eqref{spect}. We can see at glance the good
agreement of the NL approximation of the amplitude with the exact one in the
whole IR domain $0 < \om < R^{-1}$. Also the leading approximation is
qualitatively similar to the full spectrum, but its behaviour around the
transition between the flat and the decreasing regions at $\om b\sim 1$ is not
accurate. In particular, it fails to account for the (small) peak in the
spectrum around $\om b \sim 0.5$.

We analyze next the properties of the frequency spectra. We note their common
logarithmic decrease (already pointed out in~\cite{CCCV15}) in the intermediate
region $\Theta_s < \om R < 1$ ($b^{-1} < \om < R^{-1}$) which appears as a
straight line in the log-linear plot. At values of $\om R \sim \Theta_s$ the
spectra flatten out after reaching a peak and then slowly decrease towards their
ZFL limit $(\log\Theta_s^{-2}+1)/\pi$. Also clear is the common shape of the
spectra for different $\Theta_s$ in the turn-over regime $\om R\sim\Theta_s$.
In fact, by plotting the spectra against $\om b=2\om R/\Theta_s$, and by
subtracting the known ZFL, we can see that they overlap, as shown in
fig.~\ref{f:spectrum}b, where, for clarity, we limited ourselves to just two
values of $\Theta_s = 10^{-2}, 10^{-3}$. Here it is apparent that the spectrum,
starting from its finite ZFL value at $\om = 0$, increases until
$\om b\sim 0.5$ and only at larger values of the frequency it decreases.  For
the ``reduced'' spectrum, the height of the maximum above the ZFL limit is
almost independent of the (small) value of $\Theta_s$: its value is about 0.05.

\begin{figure}[htp]
  \centering
  \includegraphics[height=0.75\linewidth,angle=-90]{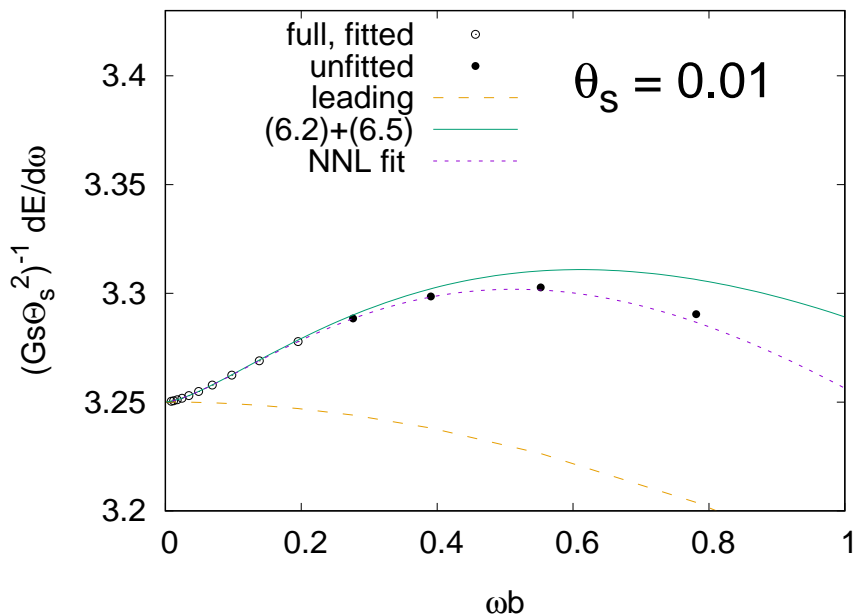}
  \caption{\it Behaviour of the spectrum in the soft limit $\om b\to0$ for
    $\Theta_s=0.01$. The full spectrum (empty and full dots) is compared with
    the one obtained from the analytic approximation~\eqref{AL}+\eqref{ANLi2}
    (solid green) and with the leading approximation~\eqref{AL} (dashed
    orange). The violet dotted line represents the function obtained by fitting
    the 10 leftmost data of the full spectrum.}
  \label{f:sciFit}
\end{figure}

This peculiar feature is due to the subleading terms of the amplitude.  In fact
the leading spectrum decreases monotonically in the whole $\omega$ range,
whereas the most relevant infrared corrections to the ZFL are positive. More
precisely, in sec.~\ref{s:sla} we found that such corrections are logarithmic
and, for the frequency spectrum, they start at $\ord{\om b\log(1/\om b)}^2$,
according to the expansion
\begin{align}
  \frac1{Gs\Theta_s^2}\frac{\dif E^\GW}{\dif\om} &\simeq \frac1{\pi} \bigg\{
    \log\frac1{\Theta_s^2}+1 \nonumber \\
    &\qquad+ \frac{(b\om)^2}{2}\left[\log^2\frac{1}{b\om}+\ord{\log\frac{1}{b\om}}
      +\ord{\Theta_s^2} \right] +\ord{\om^3\log^3\om}
  \bigg\} \;. \label{svilSpec}
\end{align}
As a consequence, the spectrum exhibits a maximum at a value of $b\om$ of order
unity. This is clearly seen by magnifying the deep IR region in linear scale.
In fig.~\ref{f:sciFit} we show the result of the full spectrum (empty and full
points) at small values of $b\om$ for $\Theta_s=10^{-2}$. By approaching
$\om\to 0$, they tend to the ZFL limit with vanishing slope, but their
behaviour is well reproduced by eq.~\eqref{svilSpec} (dotted violet curve)
which adds to the ZFL only the $[\om b\log(1/b\om)]^2$ term.

Actually, by fitting the exact spectrum with the function\\
$f(b\om) = \mathrm{ZFL} + (2\pi)^{-1}(b\om)^2[a_2\log^2(1/b\om)+a_1\log(1/b\om)+a_0]$,
(dotted violet curve) we can perfectly interpolate 10 data points within their
numerical error $\ord{10^{-5}}$, and the leading coefficient turns out to be
$a_2 = 1.001 \pm 10^{-3}$, i.e., well compatible with the theoretical
prediction. The extrapolation of $f(b\om)$ to larger values of the frequency is
able to reproduce a few more points and to reproduce their position around the
maximum (fig.~\ref{f:sciFit}).

In order to confirm the robustness of the $[\om b\log(1/b\om)]^2$ term, we have
also fitted the same data by adding to $f(b\om)$ possible next-to-leading terms
of the form $b\om[c_1\log(1/b\om)+c_0]$. By asking for a best fit we have
obtained very small values of $c_0$ and $c_1$, well compatible with zero, while
the coefficients $a_k$ at $\ord{b\om}^2$ keep their values.

To summarize, we believe that our model provides strong evidence for the
structure of the subleading coefficients in the soft limit of graviton emission
amplitudes, with terms of order $(b\om)^n\log^m(1/b\om):m\leq n$. Furthermore,
our model provides a reliable prediction for the ``dominant'' coefficients with
$m=n\leq 2$.

\section{Discussion \label{s:disc}}

In this paper we have developed our previous work on the spectrum of
gravitational waves emitted in the high-energy gravitational scattering of
massless particles at leading order in the deflection angle
$\Theta_s = \frac{4 G \sqrt{s}}{b} \equiv \frac{2 R}{b}$. This process can be
studied either at a purely classical level~\cite{GrVe14,Spirin:2015wwa} or in a
fully quantum context~\cite{CCV15,CCCV15,CC16} with the expectation that both
should agree when $\alpha_G \equiv \frac{Gs}{\hbar} \gg 1$ and the number of
produced gravitons is large.  That this is indeed the case was shown in detail
in~\cite{CCCV15} (see also~\cite{CC16}) where the second assumption was shown to
correspond to the limits $\alpha_G \Theta_s^2 \gg 1 $ and
$\frac{\hbar \om}{\sqrt{s}} \ll 1$. The overall normalization of GW spectrum
$\frac{\dif E^\GW}{\dif\om}$ is provided by its zero-frequency-limit (ZFL) and
turns out to be of order $G s \Theta_s^2 \log(\Theta_s^{-2})$.

Remarkably, the spectra obtained in this ``classical'' limit exhibit a break in
the spectrum at the characteristic ``Hawking-frequency'' scale $\om_{H} \sim
R^{-1}$. In other words the gravitational scattering process converts part of
the initial transplanckian energy into many, deeply sub-planckian, quanta (since
$\hbar/R = \sqrt{s}/\alpha_G \ll M_P$).  Below such frequency the spectrum is
almost flat, while above it decreases as $\om^{-1}$ probably up to the much
higher frequency $(\Theta_s^2 R)^{-1}$ \cite{GrVe14,CCCV15}.

In this work we have reconsidered carefully the low-frequency part of the
spectrum, $\om < 1/R$, concentrating on some small corrections at $\om < 1/b$
which, although implicitly present in the result of refs.~\cite{GrVe14,CCCV15},
had been neglected in those previous analyses. The idea of looking more closely
into this region of the spectrum was prompted by recent papers
\cite{LaSe18,SaSe18} (see also~\cite{ABV2}) in which the sub (and sub-sub)
leading corrections to soft-graviton theorems were used to compute the
corresponding sub (and sub-sub) leading corrections to the GW spectra for $\om b
\ll1$. In those papers it was pointed out that, because of the infrared
divergences of gravity in four space-time dimensions, one should expect that a
straightforward expansion in powers of $\om b$ breaks down owing to the
appearance of logarithmic enhancements. In particular, an application of the
naive recipes for computing those correction leads to infinities that can be
attributed, ultimately, to the infinite Coulomb phase characteristic of
four-dimensional physics.%
\footnote{It seems instead that the more conventional infrared divergences can
  be tamed through the usual Block-Nordsieck procedure, or, alternatively, by
  using appropriate coherent states (or the Fadeev-Kulish
  procedure~\cite{Kulish:1970ut}) without affecting the final result for the
  spectrum.}
In refs.~\cite{LaSe18,SaSe18} an improvement of the naive recipe at subleading
level was proposed, basically amounting to replacing a logarithmically diverging
time delay $\log{\tau}$ as $\tau \to \infty$ with a $\log (\om^{-1})$. This was
claimed to lead to possible observable effects, particularly on the
gravitational waveform, and also possibly of the GW spectrum for some specific
polarizations of the wave.

The advantage of the eikonal approach pursued in this paper is that it leads
directly to a singularity-free result and to an unambiguous determination of the
logarithmically enhanced contributions to the spectrum, including the
determination of the scale inside the logs. The way our approach avoids the
infinities is conceptually very simple. The infinite gravitational Coulomb
phase, as already remarked by Weinberg in 1965~\cite{We65}, comes for the
exchange of soft gravitons among the initial ``or'' the final particles (and
from singularities due to the hard-legs propagators). If the process under
consideration has just 2 hard particles in the initial state and $2+N$ in the
final state (with $N$ soft gravitons) the overall Coulomb phase for that process
is the one of the elastic $2 \to 2$ process plus the difference between the
$(2 + N)$-particle and the $2$-particle Coulomb phase. It is easy to see that
this difference is finite but contains logs. So the Coulomb divergence becomes
common to all amplitudes, factors out in impact parameter space, and cancels in
all observables; but some finite logs remain and give physical effects. We have
identified two such effects:
\begin{itemize}
\item At sub-leading order there is a correction to the ZFL of relative order
  $\om b \log(\om b)$ having interesting characteristics. It depends on the
  azimuthal angle $\phi$ of the wave vector w.r.t. the impact parameter (or
  equivalently the scattering plane) in the form of a $\pm \sin \psi$ where the
  the relation between $\phi$ and $\psi$ is given in~\eqref{angles}, and the
  sign depends on the helicity (circular polarization) of the wave. This
  interference term appears only as a $\phi$ dependent contribution to the
  polarized fluxes and cancels both in their sum and upon azimuthal averaging.
  It also disappears if one considers the more conventional $+$ and $\times$
  polarizations. All these features are in agreement with the results obtained
  in~\cite{LaSe18,SaSe18} by a very different approach.
\item At sub-sub-leading order there is instead a {\it positive} correction to
  the flux of relative order $(\om b)^2 \log^2(\om b)$, equally shared among the
  two helicities. Since this is the leading correction to the zero-frequency
  flux (with all other corrections missing the $\log^2(\om b)$ enhancement) the
  total flux must necessarily reach a maximum before falling down at higher
  $\om$. We find (both analytically and numerically) that the position of this
  maximum is at $\om b \sim 0.5$ and practically $\Theta_s$-independent.
\end{itemize}
 
It would be interesting to see how these results extend to physically more
interesting cases e.g.: i) to smaller impact parameters (i.e.\ larger deflection
angles) up to (and beyond?) the regime of inspiral; and/or, ii) to arbitrary
masses and energies of the two colliding particles.

\section{Acknowledgements}

We would like to thank the Galileo Galilei Institute for hospitality during most
of our collaboration meetings.  One of us (GV) would like to thank Andrea Addazi
and Massimo Bianchi for useful discussions about the relation between this work
and Ref.~\cite{ABV2}, Tibault Damour for discussions about the relevance of
sec.~\ref{s:iles} to the EOB program, and Ashoke Sen for informing us of his
work prior to its posting, for discussions, and for useful correspondence.


\bibliographystyle{h-physrev5}
\bibliography{all.bib}

\end{document}